\documentclass[a4paper,11pt]{article}
\pdfoutput=1 

\usepackage{jheppub,amsthm} 

\usepackage[utf8]{inputenc}

\usepackage{booktabs,multirow, quotes}

\usepackage[english]{babel}
\usepackage{amsmath,amssymb,graphicx,xcolor,alltt}
\usepackage{framed}

\newcommand{\tmop}[1]{\ensuremath{\operatorname{#1}}}

\theoremstyle{remark}

\theoremstyle{theorem}

\makeatletter
\def\@fpheader{\ }
\makeatother

\title{Local/Short-range conformal field theories from long-range perturbation theory}

\author{Junchen Rong}

\affiliation{Institut des Hautes \'Etudes Scientifiques, 91440 Bures-sur-Yvette, France}

\emailAdd{junchenrong@ihes.fr}

\abstract{We show that by imposing the conformal Wald identity, one can extract
  conformal data of the corresponding short-range/local CFT from the
  long-range perturbation theory. We first apply this to the O(N) vector
  model. We demonstrate that by properly re-sum the perturbative series, one
  gets reasonable estimations of the critical exponents of the local/short-range CFTs. We then
  apply this method to study fermionic models with four-fermion interactions.
  In 2+1 dimensions, the model has the Gross-Neveu coupling and the Thirring
  coupling. We also consider a 4+1 dimensional theory with a generalized
  Thirring coupling.}

\begin{document} 
	
	\maketitle
	
	\flushbottom

\flushbottom

\section{Introduction}

Conformal field theory (CFT) plays an important role in the theory of phase
transitions. Nearly all second-order phase transitions---whether quantum or
statistical---are described by CFTs
{\cite{cardy1996scaling,sachdev1999quantum}}. To study a CFT, certain methods,
such as Monte Carlo simulation {\cite{landau2021guide}} and conformal
bootstrap {\cite{rattazzi2008bounding,poland2019conformal,Rychkov:2023wsd}} have been
developed. In some cases
{\cite{kos2016precision,chester2020carving,chester2021bootstrapping,atanasov2022precision,erramilli2023gross}},
the bootstrap method was able to provide the highest precision results for
critical exponents. In some other cases, on the other hand, the precise
determination of the critical exponents remains unavailable, which include
various gauge theories in 2+1 dimensions
{\cite{Chester:2016wrc,Albayrak:2021xtd,He:2021sto,He:2021xvg}}. The more
traditional way to study CFTs using perturbation theories remains valuable.
These include the $4 - \epsilon$ expansion {\cite{PhysRevLett.28.240}} and the
large N expansion {\cite{Moshe:2003xn}}. Even though the corresponding
perturbative series are usually asymptotic, after proper re-summation
{\cite{bender1969anharmonic,brezin1977perturbation,brezin1978critical,kazakov1979calculation,zinn1981perturbation,le2012large}},
they give nice results for the critical exponents.

It is known that some short-range/local CFTs are in fact special points in a
conformal manifold given by long-range generalizations of the CFT. Let us take
the O(N) vector model as an example, consider the action
\[ S = \frac{\mathcal{N}^{- 1}}{2} \int d^D x d^D y \frac{\sum_i \phi_i (x)
   \phi_i (y)}{| x - y |^{D + s}} + \int d^D x  \frac{1}{8} \lambda \left(
   \sum_i \phi_i (x)^2  \right)^2 . \label{action} \]
Here $\mathcal{N}$ is a normalization factor which we will fix later. The
scaling dimension of the scalar field is $\Delta_{\phi} = \frac{D - s}{2}$. We
set $s = \frac{D + \delta}{2}$, so that $\Delta_{\phi} = \frac{D -
\delta}{4}$. With small and positive $\delta$, the interaction term is
slightly relevant. This can be viewed as a version of the analytic
regularization
{\cite{10.1007/BF02395016,gel2014generalized,bollini1964analytic,guttinger1966generalized,caianiello1973combinatorics,speer1969generalized,speer1972structure,speer1968analytic}},
that was introduced in the early days of quantum field theory. It is
well-known that the theory flows to an IR fixed point, and one can use
perturbation theory to study the corresponding long-range CFT
{\cite{fisher1972critical}}. In addition to the early works in
{\cite{fisher1972critical,sak1977low,PhysRevB.8.281,honkonen1989crossover}},
the above theory was also recently studied in small $\delta$-expansion in
{\cite{Paulos:2015jfa,Behan:2019lyd,Behan:2018hfx,Giombi:2019enr,Benedetti:2020rrq}},
in the large N limit in {\cite{Chai:2021arp}}, and using the conformal
bootstrap method in {\cite{Behan:2019lyd,Behan:2018hfx,Behan:2023ile}}. Some
Monte Carlo simulation results are also available for small N models
{\cite{picco2012critical,angelini2014relations,luijten2002boundary,Zhao:2023inj}}.

The scaling dimension of operators will depend on the parameter $\delta$. In
the $\delta \rightarrow 0$ limit, one can calculate these anomalous dimensions, such as for the mass operator $\phi^2$ and the leading spin-2
operator
$$\Delta_{\phi^2}(\delta), \quad \text { and } \quad \Delta_{T_{\mu \nu}}(\delta) \text {. }$$
The scalar operator $\phi$ will not be renormalized, so that \
\[ \Delta_{\phi} = \frac{D - \delta}{4}, \]
is valid without higher loop corrections
{\cite{fisher1972critical,sak1977low,honkonen1989crossover}}. The three loop
results of $\Delta_{\phi^2} (\delta)$ and $\Delta_{T_{\mu \nu}} (\delta)$ were
calculated in {\cite{Benedetti:2020rrq,Benedetti:2024mqx}} and in the present paper. Their
explicit forms are given in \eqref{deltaT3dindelta} and
\eqref{deltaphi23dindelta}.

It was commonly believed that the above conformal manifold contains the
short-range O(N) vector model. However, a comprehensive understanding of the
long-range to short-range crossover was only achieved recently
{\cite{Behan:2017emf,Behan:2017dwr}}. The IR CFT at the crossover is in fact
the short-range-model accompanied by a generalized free field. This explained
many long-standing puzzles of the crossover. \

In the present paper, we focus on one question, how can we extract data of
the short-range CFT from the long-range perturbation theory at small $\delta$?
Compared to the long-range CFTs in the conformal manifold, the short-range
point has a local conserved stress tensor operator, it also has a sub-sector
that satisfies the conformal Ward Identity. Reverse the logic, this means that
by imposing the corresponding conformal Ward Identity, we can extract
conformal data of the short-range models. For the O(N) vector model, one can simply impose
\[ \Delta_{T_{\mu \nu}} (\delta) = D. \]
This allows us to solve for $\delta$. Plug the result back in $\Delta_{\phi}
(\delta)$ and $\Delta_{\phi^2} (\delta)$ gives us the critical exponents of
the short range model.

\begin{figure}[h]
\begin{center}
 \resizebox{4in}{!}{\includegraphics{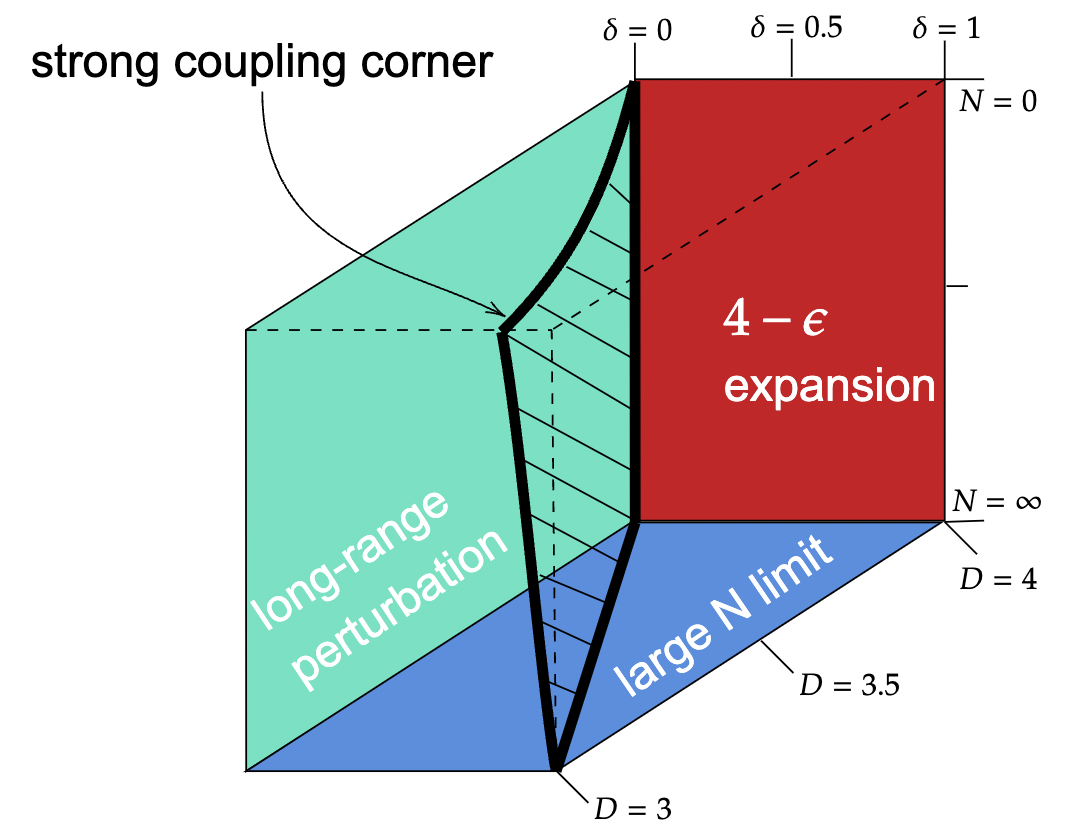}}
\end{center}
  \caption{The relation between the $4 - \epsilon$ expansion, the large N
  limit and the long range perturbation limits. \ The shaded plane is where
  the short-range models are located. The $d = 3$, $N \sim \mathcal{O} (1)$
  and $\delta \approx 1$ region is the strong coupling region.\label{threemethods}}
\end{figure}

The above procedure gives us a complementary method to study short-range/local
CFTs in addition to the $4 - \epsilon$ expansion and the large N expansion.
The relation of the three methods is summarized in Fig. \ref{threemethods}.

In this paper, we first use this approach to study the O(N) vector model. In
Section \ref{Sec:4minusepss}, we first work in $D = 4 - \epsilon$ dimensions
and show that the above procedure allows us to recover the
$\epsilon$-expansion results of the short-range models. We then work directly
in $D = 3$, after re-sum the perturbative series, we show that the above
procedure allows us to obtain reasonable results for the short-range critical
exponents, the results are summarized in Section \ref{sec:3DON}.

We then initiate the study of short-range fermionic models using long-range
perturbation theory. We note here an important motivation to develop the
long-range perturbation theory approach, namely we do not need to worry about the
complication due to the famous $\gamma^5$ problem
{\cite{leibbrandt1975introduction}} in dimensional regularization. 
When evaluating the Feynman diagram in $D = 4 - \epsilon$, one encounters
$\tmop{Tr} [\gamma^{\mu} \gamma^v \gamma^{\rho} \gamma^{\sigma}] \sim
\epsilon^{\mu \nu \rho \sigma} \tmop{Tr} [1] + \ldots$, and the Levi-civita
tensor can not be generalized to continuous dimensions. When we try to
dimensional continue to $D = 3$, on the other hand, one has to modify the
Clifford algebra so that $\tmop{Tr} [\gamma^{\mu} \gamma^v \gamma^{\rho}] \sim
\epsilon^{\mu \nu \rho} \tmop{Tr} [1] + \ldots$. At higher loops, one may need
to add back certain diagrams to recover the full symmetries of the $D = 3$
theory\footnote{This prescription was introduced in {\cite{Zerf:2017zqi}} and
named ``DREG$_3$''. Using this prescription, one successfully recovered the
supersymmetry relations for the $N = 1$ super-Ising model. See also
{\cite{Bennett:1999he}} for a treatment of Clifford algebra in $2 + \epsilon$
expansion.}.

In Section \ref{sec:3dfermion}, we study, in two-loop order, the $2 + 1$
dimensional long-range models with four-fermion couplings including the
Gross-Neveu coupling\footnote{The long-range version of the Gross-Neveu model
was first introduced in {\cite{Chai:2021wac}}.} and the Thirring coupling. In
Section \ref{sec:5dfermion}, we study, in two-loop order, a $4 + 1$
dimensional long-range fermionic model with a generalized Thirring coupling.

To calculate the Feynman integrals, we use the Mellin-Barnes method (see for
example {\cite{smirnov2013analytic}}), which can be conveniently applied using
the packages AMBRE {\cite{Gluza:2007rt}} and MB.m {\cite{Czakon:2005rk}}.

\

\section{the (scalar) O(N) vector model }\label{scalar}

We consider the long-range O(N) vector model \eqref{action}. The propagator in
real space is
\[ G_{i j} (x - y) = (H_s H_{- s})^{- 1} \frac{\mathcal{N}}{| x - y |^{D -
   s}} \delta_{i j}, \qquad \tmop{with} \quad H_s = \frac{2^s
   \pi^{\frac{d}{2}} \Gamma \left( \frac{s}{2} \right)}{\Gamma \left( \frac{d
   - s}{2} \right)}, \]
which satisfies the long range equation of motion,
\[ \frac{1}{\mathcal{N}} \int d^D y \frac{\delta_{i j}}{| x - y |^{D + s}}
   G_{j k} (y - z) = \delta_{i k} \delta^{(D)} (x - z) . \]
We have used the equation,
\[ \int d^D y \frac{1}{| x - y |^{D + s}} \frac{1}{| y - z |^{D - s}} = H_s
   H_{- s} \delta^{(D)} (x - z) . \]
\[ \  \]
Using the Fourier transformation, one can show that
\[ \int d^D x \frac{1}{r^{D - s}} e^{- i p \cdot x} = H_s \frac{1}{| p |^s},
   \qquad \frac{1}{(2 \pi)^D} \int d^D p   \frac{1}{| p |^s} e^{i p \cdot x} =
   (H_s)^{- 1} \frac{1}{r^{D - s}} . \]
We take $\mathcal{N} = H_{- s}$, so that the propagator in the momentum space
is given by,
\[ G (p)_{i j} = \frac{\delta_{i j}}{(p^2 + \mu^2)^{s / 2}} . \]
We
work in Euclidean signature, so that $p^2 = \sum^3_{i = 1} p_i^2$. Notice that we have introduced the IR regulator $\mu$ for the propagator. It is
important to introduce this regulator to remove the IR singularity, so that the
divergence of the Feynman integral contains UV divergence only. The Feynman
rule for the four-point vertex is
\[ V_{i j k l} = \lambda (\delta_{i j} \delta_{k l} + \delta_{i l} \delta_{j
   k} + \delta_{i k} \delta_{j l}) . \]
We take $s = \frac{D + \delta}{2}$, so that the scaling dimension of the
scalar operator is
\[ \Delta_{\phi} = \frac{D - \delta}{4} . \]
In the $\delta \rightarrow 0_+$ limit, the $\lambda \phi^4$ coupling term is
slightly relevant. We therefore take $\delta$ to be the perturbation
parameter. The short-range models are located in the $\delta \approx 1$
region.

We collect here the perturbative calculation results of the theory, the beta
function was calculated in earlier papers, in our convention (which is the
same convention as in {\cite{Benedetti:2020rrq}}),
\[ \beta_g = - \delta g + \frac{1}{2} \delta g^2  (N + 8) D_0 + \delta g^3  (5
   N + 22)  (D_0^2 - 2 S_0), \]
The order $g^4$ result of the beta function is only known up to an infinite sum, interested reader shall refer to \cite{Benedetti:2024mqx}.
We use $g$ to denote the renormalized coupling constant, and $\lambda$ to
denote the bare coupling (see \ {\cite{Benedetti:2020rrq}}). The constants
$D_0$ and $S_0 $ were calculated in {\cite{Benedetti:2020rrq}},
\begin{equation}
  D_0 = \frac{2^{- D} \pi^{- \frac{D}{2}} \Gamma \left( \frac{\delta}{2}
  \right)}{\Gamma \left( \frac{D + \delta}{2} \right)}, \qquad S_0 =
  \frac{2^{1 - 2 D} \pi^{- D}}{\delta^2 \Gamma \left( \frac{D}{2} \right)^2} -
  \frac{(4 \pi)^{- D}  \left( \psi^{(0)} \left( \frac{D}{2} \right) + 2
  \psi^{(0)} \left( \frac{D}{4} \right) + 3 \gamma \right)}{\delta \Gamma
  \left( \frac{D}{2} \right)^2} . \label{DScoupling}
\end{equation}
The anomalous dimension of the leading spin-2 operator is given by
\begin{equation}
  \Delta_{T^{\mu \nu}} = 2 \Delta_{\phi} + 2 - 3 \delta g^2  (N + 2) S_2 +
  \frac{3}{4} \delta g^3  (N^2 + 10 N + 16)  (- 4 D_0 S_2 + 4 I_1 + I_2),
  \label{spin2ing}
\end{equation}
with $S_2$ defined in \eqref{s2constant}, $I_1$ and $I_2$ defined in
\eqref{i2constant}.
At the limit $\delta \rightarrow 0$, we have 
\begin{align}
    T_{\mu\nu}=&(\phi^i\partial_{\mu}\partial_{\nu}\phi^i-\frac{1}{D}\delta_{\mu\nu}\phi^i\partial^2\phi^i)\nonumber\\
    &-\frac{\Delta_{\phi}+1}{\Delta_{\phi}}(\partial_{\mu}\phi^i\partial_{\nu}\phi^i-\frac{1}{D}\partial_{\mu}\phi^i\partial^{\mu}\phi^i).
\end{align}
The coefficients between the two terms ensure that this operator is a conformal primary.
The anomalous dimension of the leading spin-1 operator (in the adjoint irrep
of O(N)) is
\begin{equation}
  \Delta_{J^{\mu}} = 2 \Delta_{\phi} + 1 - \delta g^2 (N + 2) S_1 +
  \frac{1}{4} \delta g^3 (N + 2)  (- 4 D_0 S_1 + 24 I_1' + 3 (N + 4) I_2') .
  \label{spin1ing}
\end{equation}
with $S_1$ defined in \eqref{s1constant}, \ $I_1'$ defined in
\eqref{i1pconstant} and $I_2'$ defined in \eqref{i2pconstant}.

From the beta function, we can solve the coupling constant at the fixed point
${g_{\star}}_{}$ in terms of the perturbation parameter $\delta$. From which
we get the scaling dimension of the spin-2 operator at the O(N) long-range
fixed points at three-loop,
\begin{eqnarray}
  \Delta_{T^{\mu \nu}} & = & \frac{D + 4}{2} - \frac{\delta}{2} - \frac{12
  \delta^2  (N + 2)}{D (D + 2)  (N + 8)^2} \nonumber\\
  &  & + \frac{2 \delta^3  (N + 2)}{D^2  (D + 2)  (N + 8)^4} \bigg[ \frac{(D
  (3 D - 20) - 64) (N + 8)^2}{D + 4}  \nonumber \\&&+ D (N - 4) (7 N + 20) \left( \psi^{(0)}
  \left( \frac{D}{2} \right) + \gamma \right) - 2 D (N - 4) (7 N + 20) \left(
  \psi^{(0)} \left( \frac{D}{4} \right) + \gamma \right) \bigg] .\nonumber\\
  \label{deltaT3dindelta} 
\end{eqnarray}
(The two loop result of $\Delta_{T^{\mu \nu}}$ was previously calculated in
{\cite{Behan:2019lyd,Behan:2018hfx,Behan:2023ile}}). 
Similarly using the
$\Delta_J$ calculated in \eqref{spin1ing}, we get the scaling dimension of the
leading spin-1 operator,
\begin{eqnarray*}
  \Delta_{J^{\mu}} & = & \frac{D + 2}{2} - \frac{\delta}{2} - \frac{2 \delta^2
  (N + 2)}{D (N + 8)^2} +\\
  &  & + \frac{\delta^3  (N + 2) }{D^2  (N + 8)^4} \bigg( D ((N - 16) N - 48)
  \left( \psi^{(0)} \left( \frac{D}{2} \right) + \gamma \right) \nonumber\\&&- 2 D ((N -
  16) N - 48) \left( \psi^{(0)} \left( \frac{D}{4} \right) + \gamma \bigg) -
  32 (N + 8) \right) .
\end{eqnarray*}
For completeness, we also note down the scaling dimension of the $\phi^2$
operator which are known at three-loop in {\cite{Benedetti:2020rrq}}. At two loop order, it is
\begin{eqnarray} \label{deltaphi23dindelta}
  \nu^{- 1} & = & \frac{D}{2} + \delta \left( \frac{6}{N + 8} - \frac{1}{2}
  \right) + \frac{\delta^2  (N + 2)  (7 N + 20)  \left( - \psi^{(0)} \left(
  \frac{D}{2} \right) + 2 \psi^{(0)} \left( \frac{D}{4} \right) + \gamma
  \right)}{(N + 8)^3} 
\end{eqnarray}
The $\delta^3$ order result are known up to an infinite sum \cite{Benedetti:2024mqx}. In $D=3$, we have 
\begin{eqnarray}
 \nu^{- 1} \big|_{ \mathcal O(\delta^3)}&=\quad & 0.3536 \delta^3,\quad {\rm for }\quad N=1,\nonumber \\
 &&0.3608\delta^3,\quad {\rm for } \quad N=2,\nonumber \\
  &&0.3556\delta^3, \quad {\rm for } \quad N=3.\nonumber 
\end{eqnarray}
The critical exponent $\omega$, calculated in {\cite{Benedetti:2020rrq}}, is
\begin{eqnarray} \omega &&= \delta +\delta^2  \frac{2  (5 N + 22)  \left( - \psi^{(0)} \left(
   \frac{D}{2} \right) + 2 \psi^{(0)} \left( \frac{D}{4} \right) + \gamma
   \right)}{(N + 8)^2} \nonumber.
\end{eqnarray}
Similarly, the $\delta^3$ results are known up to an infinite sum \cite{Benedetti:2024mqx}. In $D=3$, we have 
\begin{eqnarray}
 \omega \big|_{ \mathcal O(\delta^3)}&=\quad & 4.021 \delta^3,\quad {\rm for }\quad N=1,\nonumber \\
 &&3.625\delta^3,\quad {\rm for } \quad N=2,\nonumber \\
  &&3.292\delta^3, \quad {\rm for } \quad N=3.\nonumber 
\end{eqnarray}
We will also need the scaling dimension of the so called ``double twist'' operator $[\phi\phi]_{n,l}$, whose two-loop anomalous dimension were calculated in \cite{Behan:2023ile},
\begin{align}
\Delta_{[\phi\phi]_{n,l}}=\frac{D+4 n+2l}{2}-\frac{\delta}{2}-\delta^2\frac{N+2}{N+8}\frac{(-1)^n}{n!}\frac{3\Gamma(\delta/2)\Gamma(n+l)\Gamma(D/2+n)}{\Gamma(D/2+2n+l)\Gamma(D/2-n)}
\end{align}
The operator with $n=1, l=0$ is an operator with fairly low anomalous dimension and maybe affect the critical exponents $\omega$. At the $\delta\rightarrow 0$ limit, the operator $[\phi\phi]_{1,0}$ is given by 
 \begin{align}\label{doubletrace0}
\frac{2\Delta_{\phi}}{2\Delta_{\phi}-(D-2)}(\partial^2\phi_i)\phi_i-\partial_{\mu}\phi_i\partial^{\mu}\phi_i.
 \end{align}
\subsection{$4 - \epsilon$ expansion, a consistency
check}\label{Sec:4minusepss}

With the results in the previous section, we now show that one can recover the
short-range CFT data. We make an ansatz $\delta = a_1 \epsilon + a_2
\epsilon^2 + \ldots .$ and solve the following equations
\[ \Delta_{T^{\mu \nu}} (\epsilon, \delta) = 4 - \epsilon, \quad \tmop{or}
   \quad \Delta_{J^{\mu}} (\epsilon, \delta) = 3 - \epsilon, \]
up to $\epsilon^3$ order. For both equations, we get the same solution,
\begin{equation}
  \delta = 2 \epsilon - \frac{4 (N + 2) \epsilon^2}{(N + 8)^2} + \frac{2 (N^3
  - 54 N^2 - 384 N - 544) \epsilon^3}{(N + 8)^4} + \ldots, \label{deltaeps}
\end{equation}
One recognizes that $\Delta_{\phi} = \frac{d - \delta}{4}$ agrees with the
calculation in $4 - \epsilon$ perturbation theory
{\cite{kleinert2001critical}}. Substitute \eqref{deltaeps} in \eqref{deltaphi23dindelta}, we again recover the
result of short-range models in $4 - \epsilon,$
\begin{eqnarray*}
  \nu^{- 1} & = & 2 + \frac{(- N - 2) \epsilon}{N + 8} + \frac{(- N - 2)  (13
  N + 44) \epsilon^2}{2 (N + 8)^3}  \nonumber  .
\end{eqnarray*}
Notice these results do not generalize to the critical exponents $\omega =
\frac{d \beta (g)}{d g}$. Namely, plug \eqref{deltaeps} into $\omega$ does not
recover the $\omega$ critical exponents in $4 - \epsilon$ for the short-range
O(N) vector model. In the traditional short-range $4 - \epsilon$ expansion, we
have three almost marginal operators,
\begin{equation}
  \partial_{\mu} \phi \partial_{\mu} \phi, \quad \phi \partial^2 \phi, \quad
  \tmop{and} \quad \phi^4, \label{shortrange}
\end{equation}
One combination of the three operators is the level two descendant of
$\phi^2$. Another combination becomes a null operator and decouples from the
spectrum due to the equation of motion $\partial^2 \phi \sim g \phi^3$. We are
then left with a single operator which is a conformal primary, whose scaling
dimension gives us the critical exponents $\omega$. For long-range model, the long-range 
equation of motion does not cause any combinations of these three operators to become null. 
We have two conformal primaries $\phi^4$ and $[\phi\phi]_{1,0}$ as defined in \eqref{doubletrace0}.
They are almost marginal and will mix when $D=4-\epsilon$.
To get
$\omega_{\tmop{short}}$ we have to resolve such a mixing, which we leave for
future work.
We will see that in $D=3$ such a mixing does not happen.

\subsection{re-summation of the $D = 3$ perturbative series }\label{sec:3DON}

Even though our perturbative series is known only to three-loop order, it is,
however, interesting to think about how we can re-summation this series. For a
nice review of the re-summation theory in quantum field theory, see
{\cite{kleinert2001critical}}. For an asymptotic series with zero radius of
convergence
\[ A' (g) = \sum_{k = 0}^{\infty} f_k g^k, \]
one can define the Borel-Leroy series to be
\[ B^{\beta} (g) = \sum_{k = 0}^{\infty} \frac{f_k}{\Gamma (k + \beta + 1)}
   g^k . \label{borel} \]
The inverse transformation is defined as
\[ A  (g) = \int d t t^{\beta} e^{- t} B^{\beta} (g t) . \]
For a perturbative series, the Borel-Leroy procedure leaves the series
invariant. This may seem useless. However, in many cases, $B^{\beta}
(g)$ is a convergent series. If one can sum the $B^{\beta} (g)$
series, the inverse Borel-Leroy transformation will give us an analytic
function $A (g) .$

Suppose the singularity of $B^{\beta} (g)$ that is closest to the origin is
located at $g = - 1 / \alpha$, with a branch cut given by
\[ B^{\beta} (g) \sim \frac{\gamma}{(1 + \alpha g)^{\beta + 1}},
\]
The inverse Borel-Leroy transformation tells us the $f_k$ asymptotes to
\[ f_k = \gamma k^{\beta} k! (- \alpha)^k, \qquad k \rightarrow \infty .
   \label{largeasym} \]
For a field theory, large $k$ behavior is usually controlled by the saddle
points of the path integral, in other words, instanton solutions of the
classical equation of motion. The constant $\alpha$ is the action of the
leading instanton solution.

If the function $B^{L, \beta} (g) = \sum_{k = 0}^L \frac{f_k}{\Gamma (k
+ \beta + 1)} g^k$ is only known to a limited number of terms, one can
perform a conformal mapping
\[ w (g) = \frac{\sqrt{\alpha g + 1} - 1}{\sqrt{\alpha g + 1} +
   1}, \]
to map the branch cut onto the unit circle. Re-expansion $B^{\beta} (g)$
using the new variable $w$, the truncated series
\[ B^{' (w) \beta} (g) \approx \sum_{k = 0}^L W_k w^k (g) \]
are expected to converge faster. Clearly the above approach is only possible
if we know the instanton solutions.

In addition to the conformal mapping technique, another way to approximate the
Borel function is to use the Pade approximation. This is the so-called
Pade-Borel re-summation method. To do this, one imposes that
\[ \tmop{PB}^{L, \beta} (g) = \frac{\sum^M_i a_i g^i}{\sum^N_i b_i
   g^i} + \mathcal{O} (g^{L + 1}) = \sum_{k = 0}^L
   \frac{f_k}{\Gamma (k + \beta + 1)} g^k, \quad \tmop{with} \quad M + N
   = L. \]
In other words, we require $\tmop{PB}^{L, \beta} (g) = B^{L, \beta}
(g)$ up to order $g^{L + 1}$. The freedom in choosing $(M, N)$ is
an ambiguity. This ambiguity can be partially resolved by requiring the
singularity of \ $\tmop{PB}^{L, \beta} (g)$ to resemble the singularity
structure of $B^{\beta} (g)$. In our case, this means that the poles of
\ $\tmop{PB}^{L, \beta} (g)$ should lie close to the negative
$g$-axis.

We will use both the conformal mapping method and the Borel-Pade re-summation
method in our analysis. We now first discuss the instanton solutions. In the
$\delta \rightarrow 0$ limit, the solution at $g < 0$ is given by the solution
of the following integral equation,
\[ \frac{1}{\mathcal{N}} \int d^D y \frac{1}{| x - y |^{D + s}} \phi_i (y) = -
   g \frac{1}{2} \phi_i (x) \sum_j \phi_j (x)^2 . \]
The equation of motion has a solution given by
\[ \overrightarrow{\phi_c} (x) = \hat{n} \times \sqrt{2}  \left( -
   \frac{\pi^{- \frac{D}{2}} g \mathcal{N} \Gamma \left( \frac{D - s}{2}
   \right)}{\Gamma \left( - \frac{s}{2} \right)} \right)^{\frac{s - D}{2 s}}
   \left( \frac{a}{a^2 + |x - x_0 |^2} \right)^{\frac{D - s}{2}}, \qquad
   \tmop{with} \quad s = \frac{D}{2} . \]  
This is a simple generalization of the $N = 1$ instanton solution first written down by Elliott Lieb in \cite{LiebSharp}.
This integral is closely related to fractional Laplacian, and has been
extensively studied by mathematicians (such as in
{\cite{chen2006classification}})\footnote{We thank Kihyun Kim for pointing out
to us the long-range instanton in the $N = 1$ model.}. The moduli space for the
instanton solution is parametrized by the vector $\hat{n}$ satisfying $\hat{n}
\cdot \hat{n} = 1$ and the parameter $a$, which controls the size of the
instanton. The action of the solution is
\[ S = - \frac{1}{8} g \int d^D x \left( \sum_i \phi_{c, i} (x)^2 
   \right)^2 =-\frac{\pi ^{\frac{D+1}{2}} \Gamma \left(\frac{3 D}{4}\right)^2}{g  \Gamma \left(\frac{D}{4}\right)^2 \Gamma \left(\frac{D+1}{2}\right)} .
\]
When $D = 3$, $S = - \frac{\pi^2 \Gamma \left( \frac{9}{4} \right)^2}{g
\Gamma \left( \frac{3}{4} \right)^2} = - \frac{1}{\alpha g}$.  In practice, we directly resum the perturbative series in $\delta$, so we will need the critical coupling $g_{*}=\frac{2^D\pi^{D/2}\Gamma(D/2)}{N+8}$.
\

The constant $\gamma$ and $\beta$ in \eqref{largeasym} depend on the (massive
and massless) fluctuation around the instanton. 
For the short-range
model, one needs to solve a Schrodinger problem with the potential given by
the instanton. This can be done analytically (see, for example,
{\cite{Zinn-Justin:1980oco}}). It will be interesting to study the analogous
problem for long-range models, we leave this for future work. For simplicity,
we will take
\[ \beta = 0. \]
One will see that this already gives us reasonable results.

\begin{figure}[h]
\begin{center}
 \resizebox{0.6\columnwidth}{!}{\includegraphics{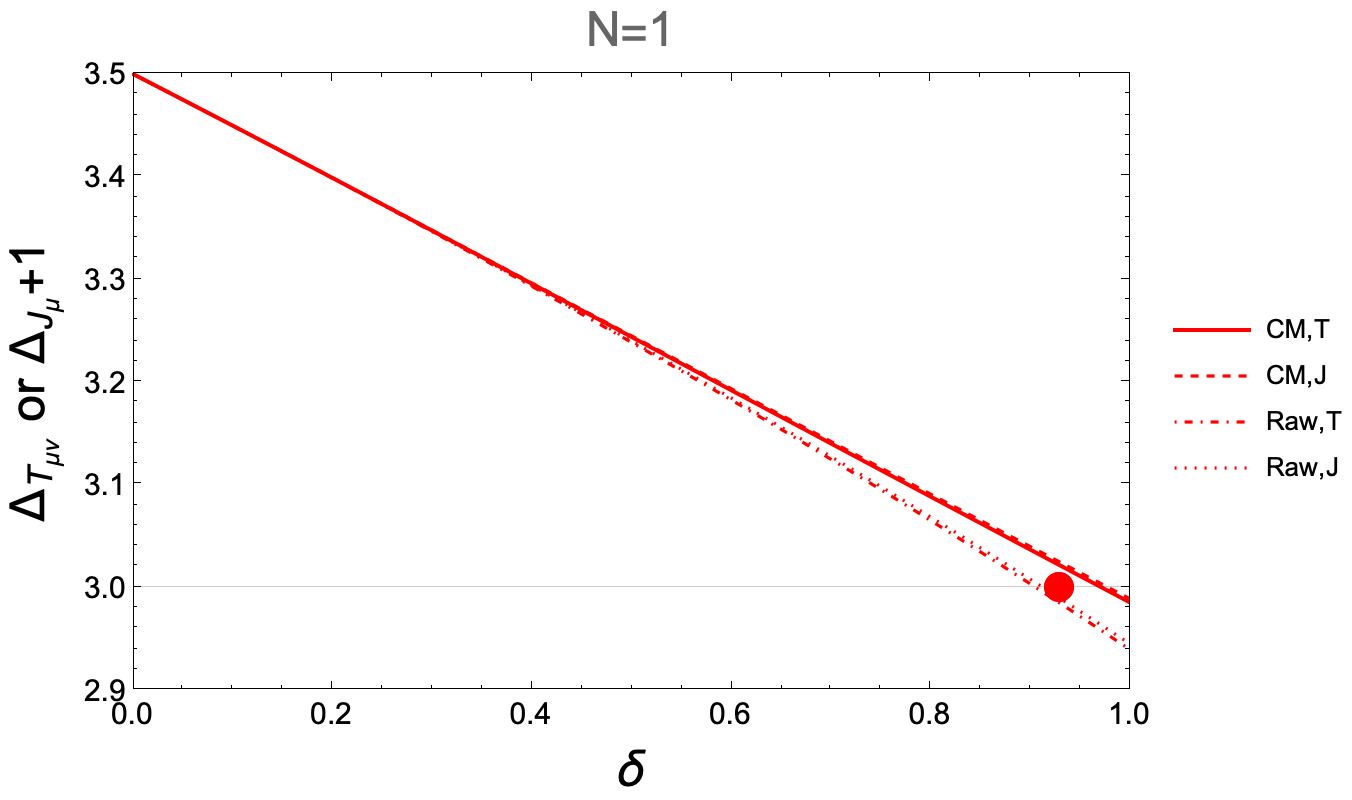}}
  
  \resizebox{0.6\columnwidth}{!}{\includegraphics{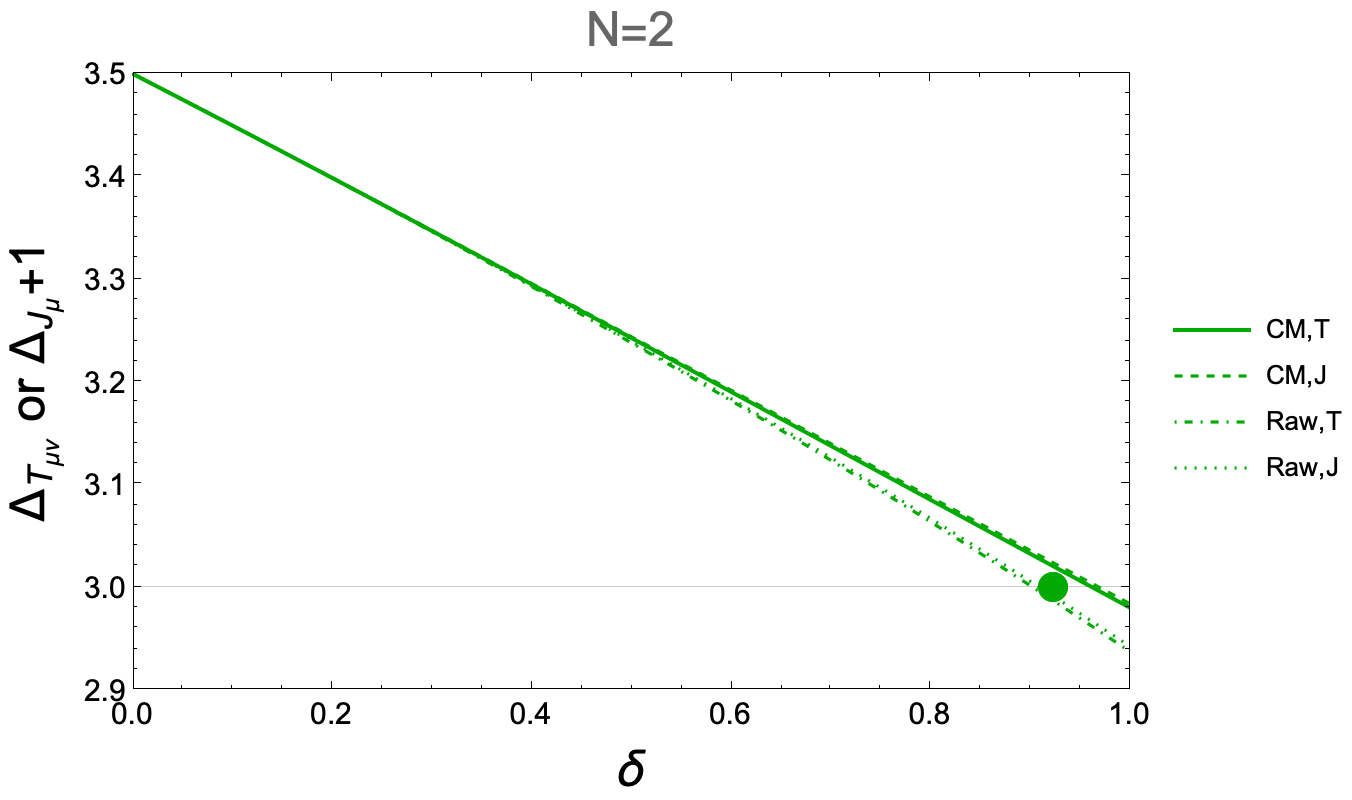}}
  
  \resizebox{0.6\columnwidth}{!}{\includegraphics{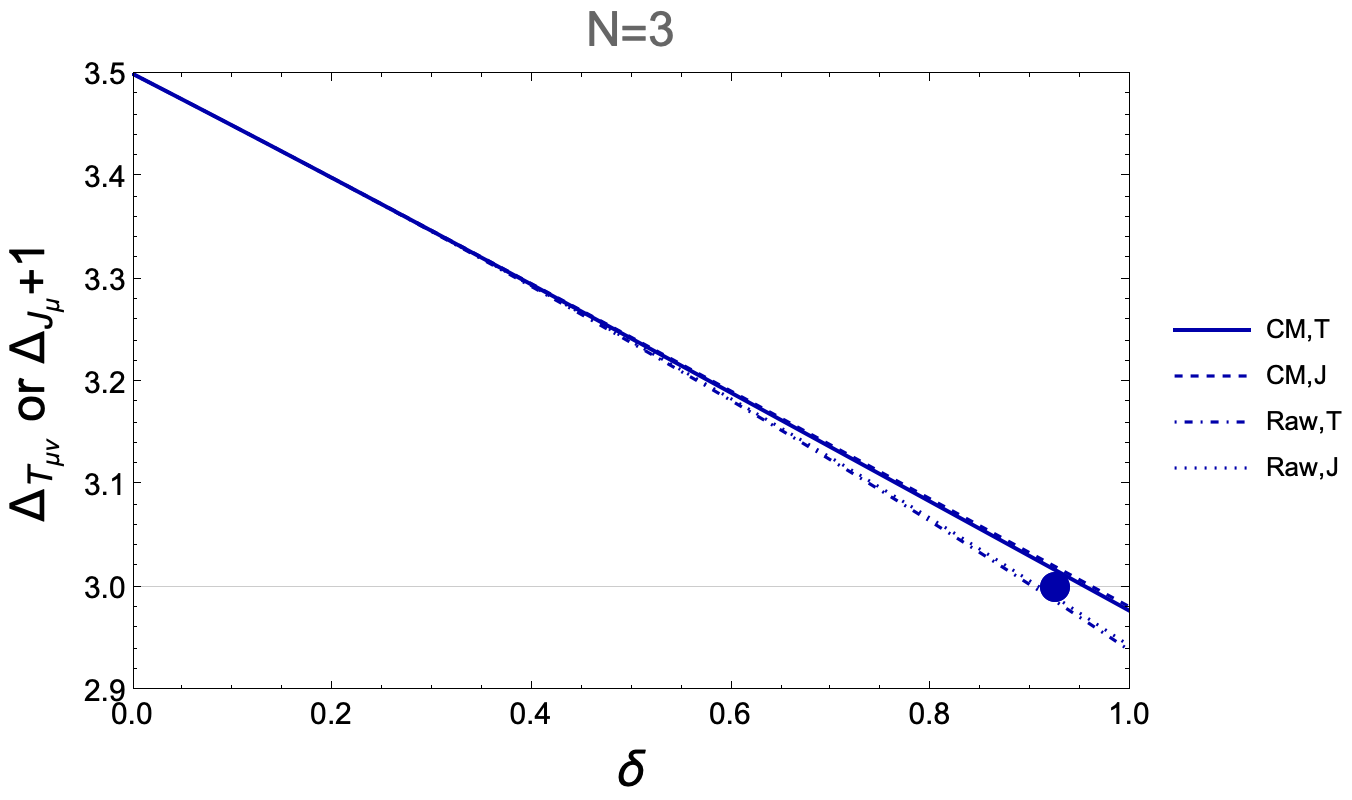}}
\end{center}
   \caption{Re-summation of $\Delta_T$ and $\Delta_J + 1$ for $N = 1, 2, 3$.
  The solid line is the result of the re-summation of $\Delta_T$ using
  conformal mapping technique. The dashed line is the result of the
  re-summation of $\Delta_J + 1$ using conformal mapping technique. The
  dash-dotted line is $\Delta_T$ without re-summation. The dotted line is
  $\Delta_J + 1$ without re-summation. We also tried to use the Pade-Borel
  re-summation method. It turns out the Pade approximations of the
  corresponding Borel functions have poles with positive real parts, we
  therefore discard all cases. The dots indicate the results from conformal bootstrap \cite{kos2016precision,chester2020carving,chester2021bootstrapping}.\label{deltaTJ} }
\end{figure}

\begin{figure}[h]
\begin{center}
  \resizebox{0.6\columnwidth}{!}{\includegraphics{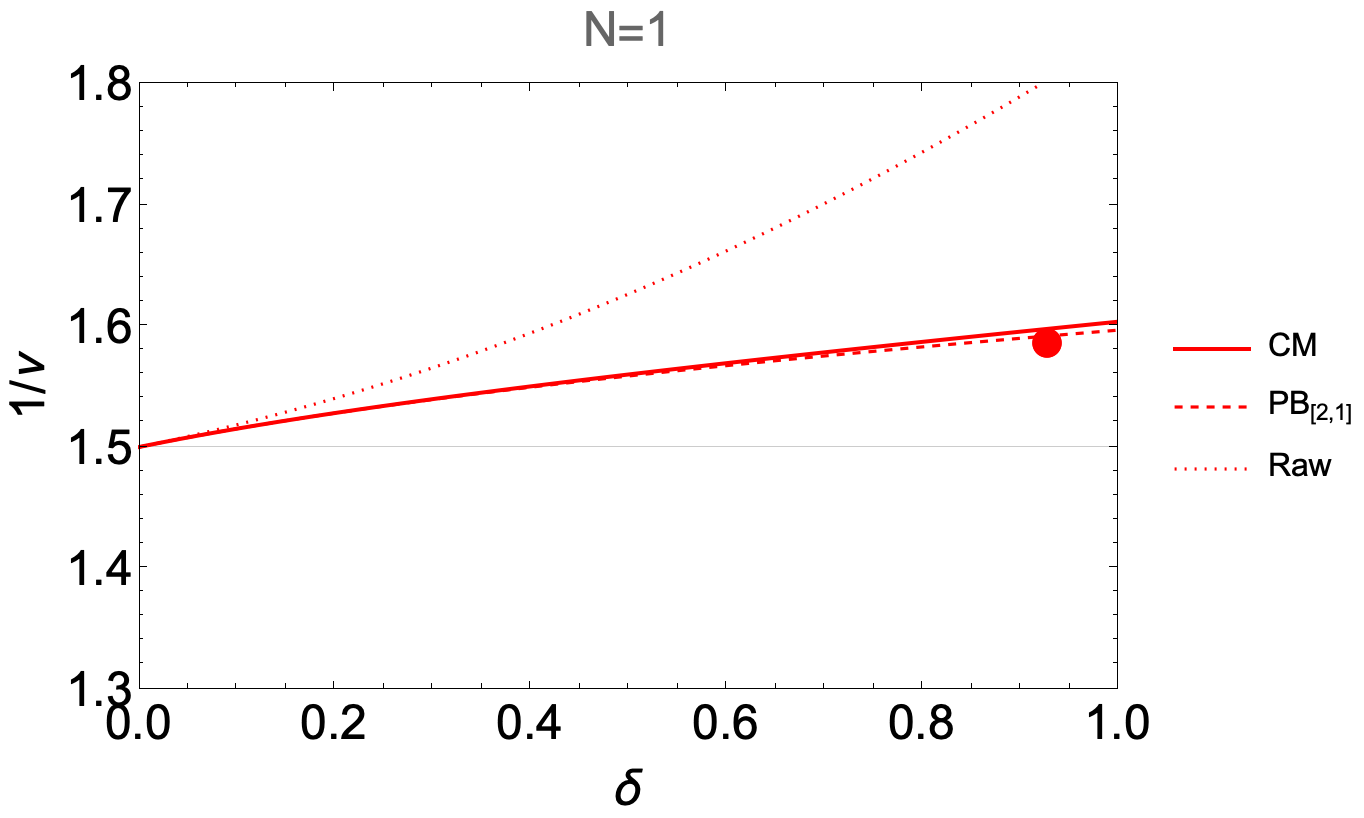}}
  
  \resizebox{0.6\columnwidth}{!}{\includegraphics{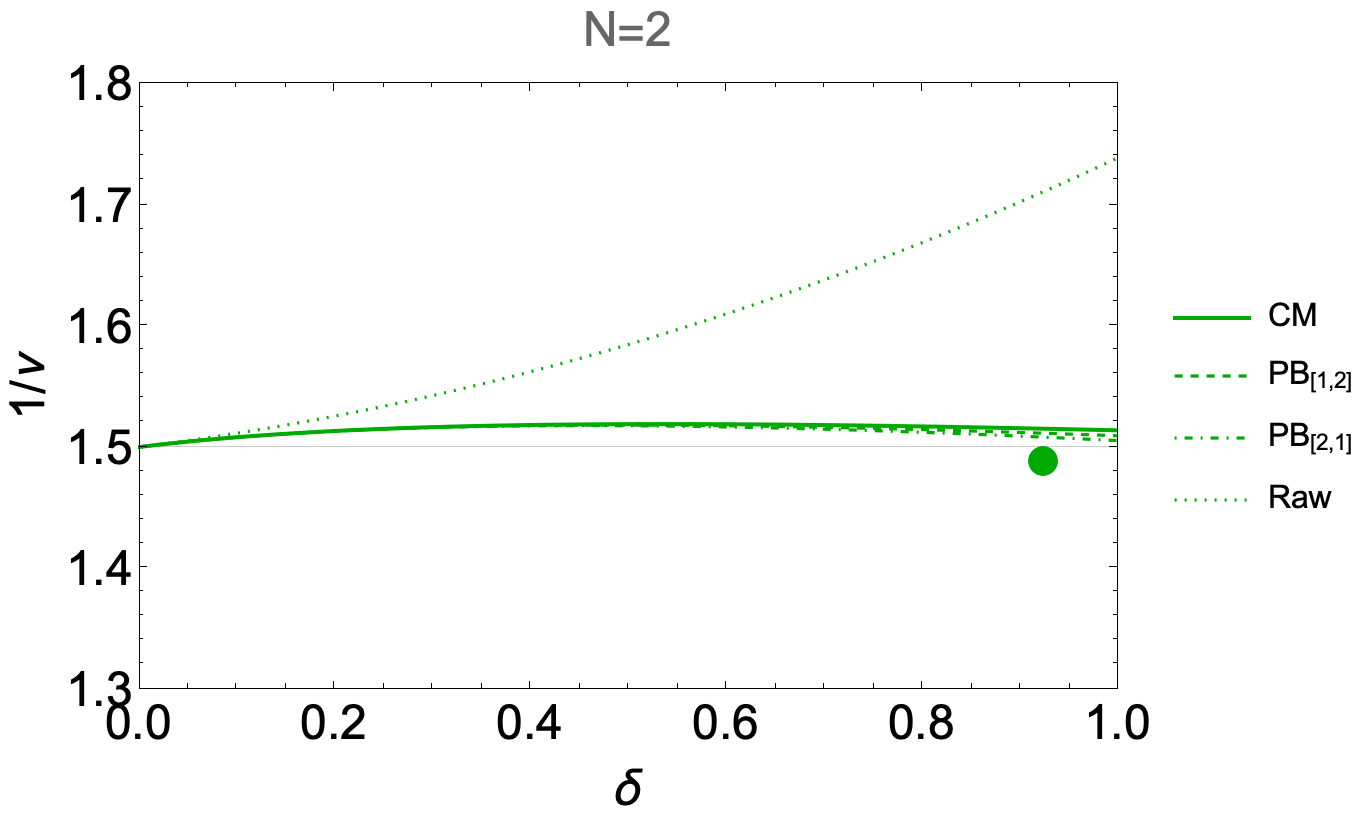}}
  
  \resizebox{0.6\columnwidth}{!}{\includegraphics{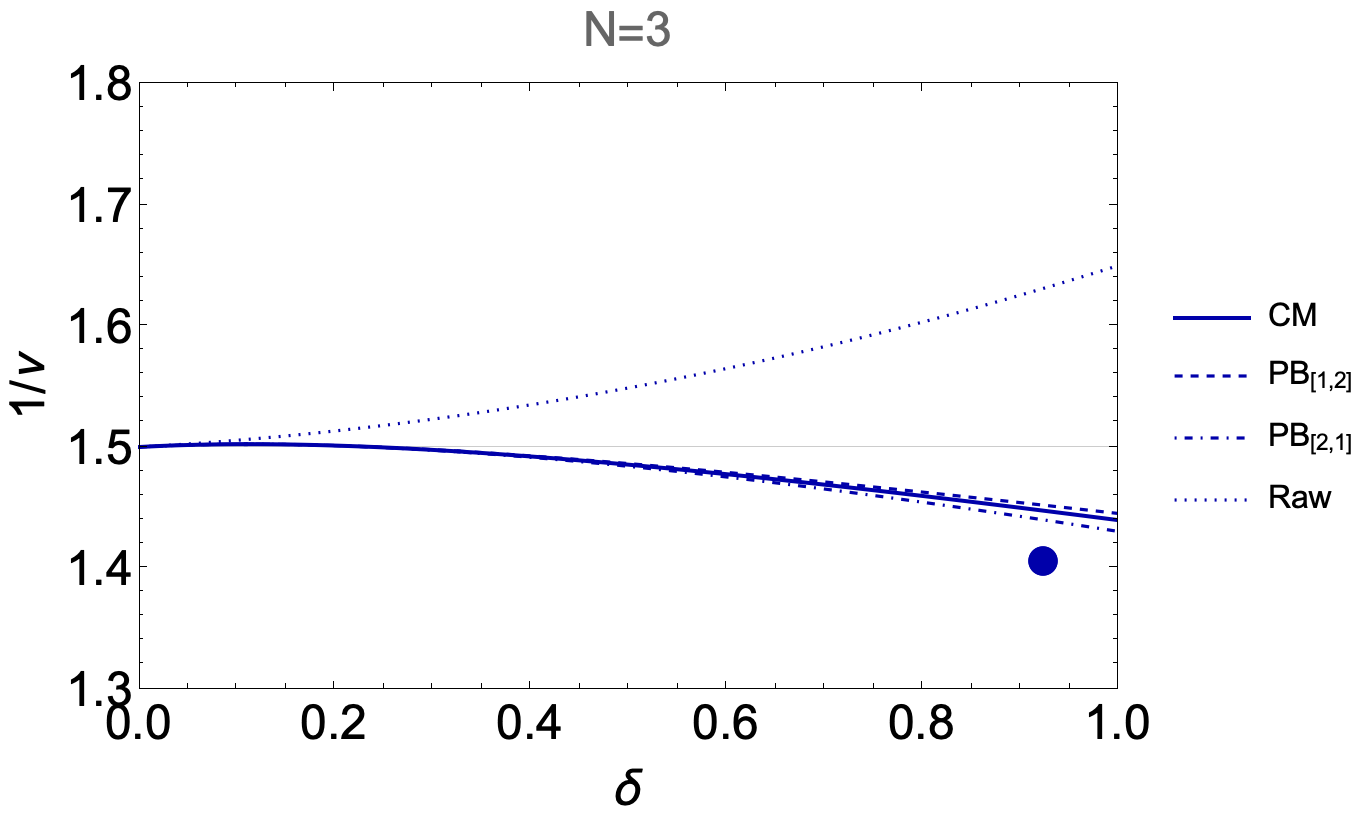}}
\end{center}
  \caption{Re-summation of $\nu^{- 1} = 3 - \Delta_{\phi^2} $ for $N = 1, 2,
  3$. The solid line is the result of the re-summation using conformal mapping
  technique. The dashed line is the result using Pade$_{[1, 2]}$-Borel method.
  The dash-dotted line are the results using Pade$_{[2, 1]}$-Borel method. \
  The dotted line is the result without re-summation.\label{nuplot} For the
  Pade$_{[0, 3]}$ result, the Pade approximations of the corresponding Borel
  function has poles with positive real parts, we discard this case. The dots indicate the results from conformal bootstrap \cite{kos2016precision,chester2020carving,chester2021bootstrapping}.} 
\end{figure}

\begin{figure}[h]
\begin{center}
  \resizebox{0.6\columnwidth}{!}{\includegraphics{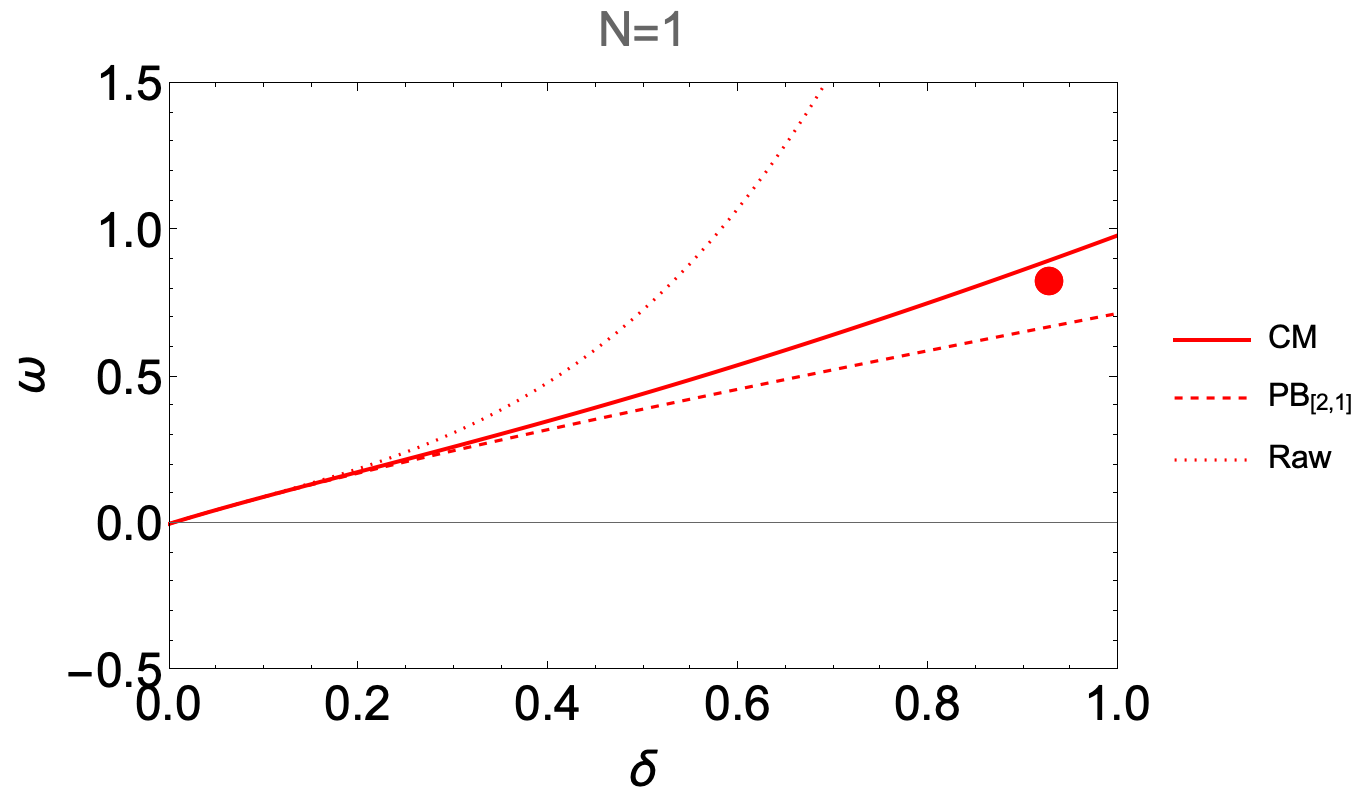}}
  
  \resizebox{0.6\columnwidth}{!}{\includegraphics{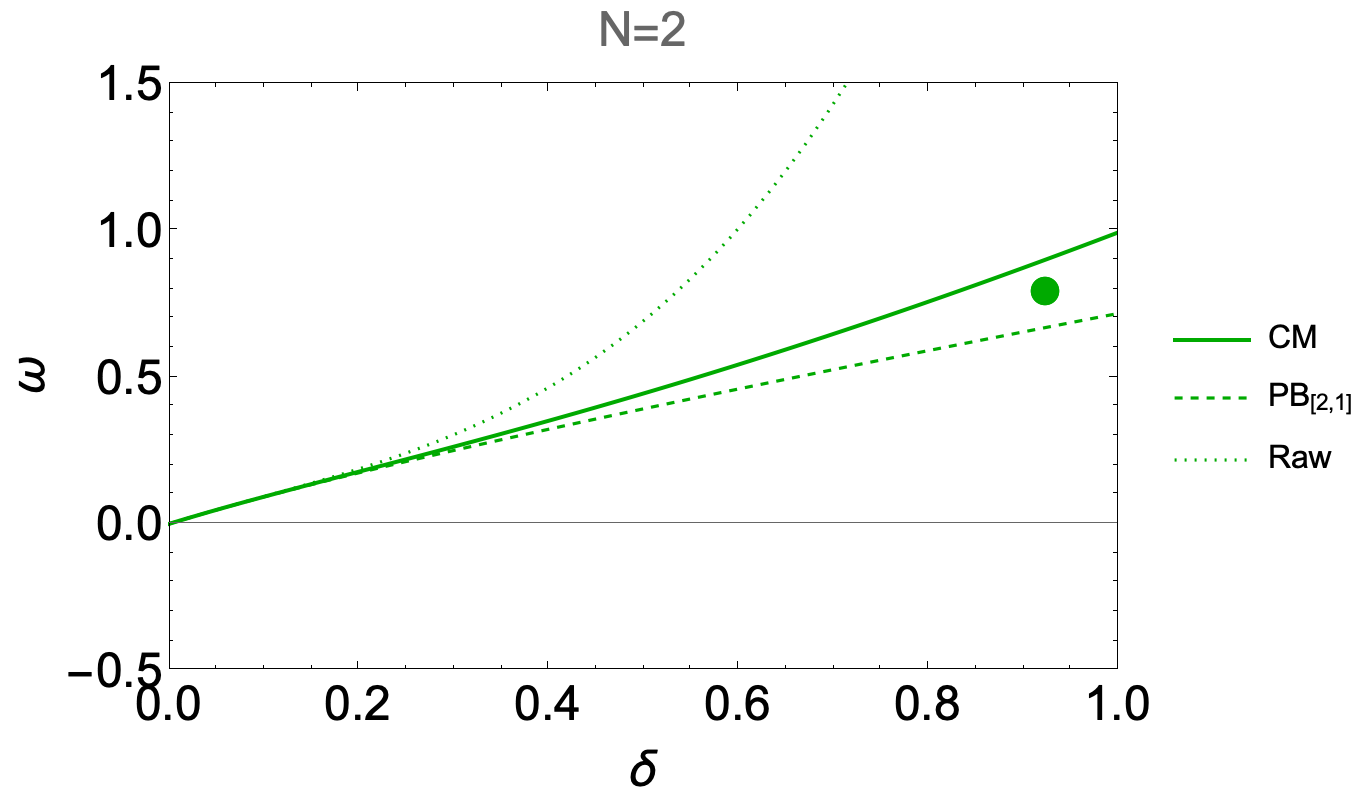}}
  
  \resizebox{0.6\columnwidth}{!}{\includegraphics{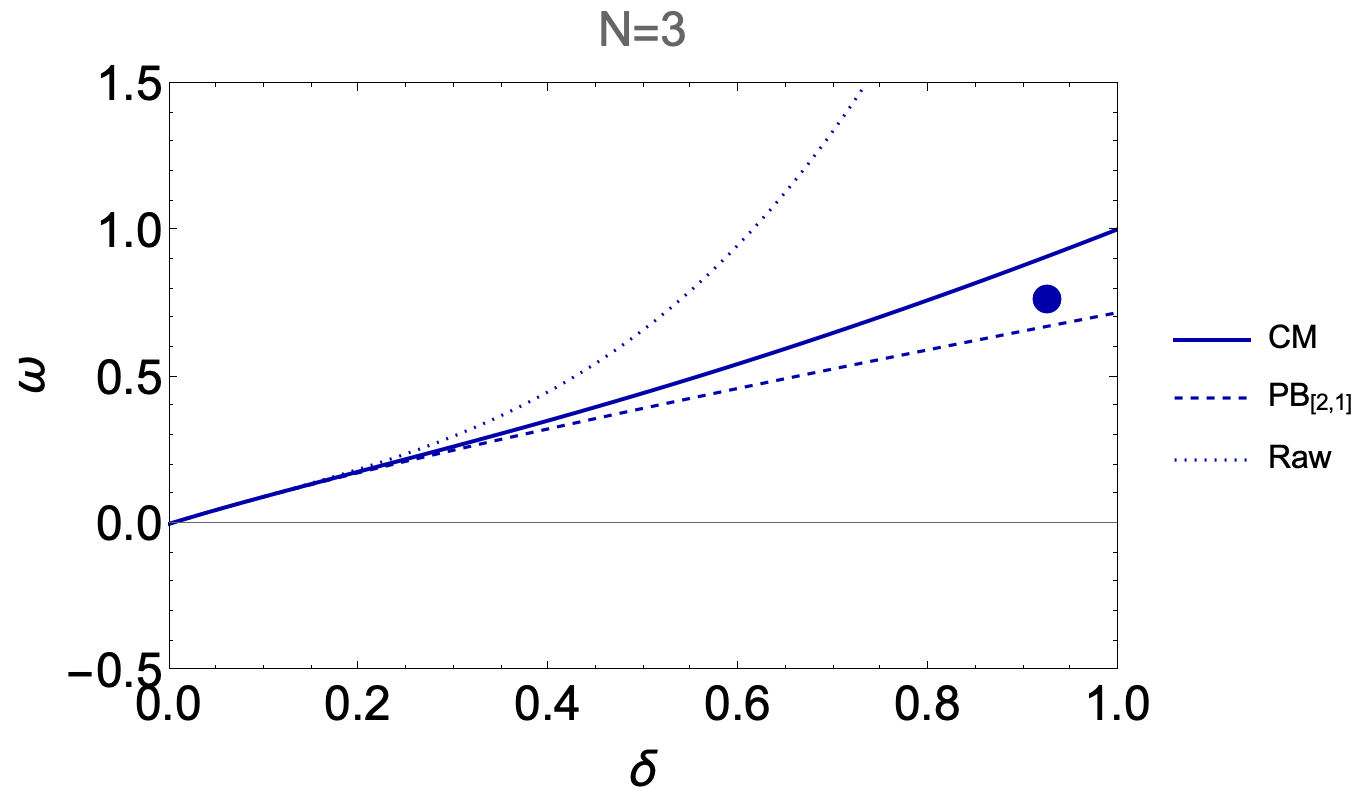}}
  
\end{center}
  \caption{Re-summation of $\omega$ for $N = 1, 2, 3$. Here $\omega=\partial{d\beta(g)}{dg}|_{g=g_*}$. The scaling dimension of the operator $\phi^4$ is given by $\Delta_{\phi^4}=3+\omega$. The solid line is the
  result of the re-summation using conformal mapping technique. The dashed
  line is the result using Pade$_{[2, 1]}$-Borel method. \ The dotted line is
  the result without re-summation. Sometimes, the Pade approximations of the
  corresponding Borel functions have poles with positive real parts, we
  discard these cases. The dots indicate the results from conformal bootstrap \cite{kos2016precision,chester2020carving,chester2021bootstrapping}.\label{omegaplot} }
\end{figure}

\begin{figure}[h]
\begin{center}
  \resizebox{0.6\columnwidth}{!}{\includegraphics{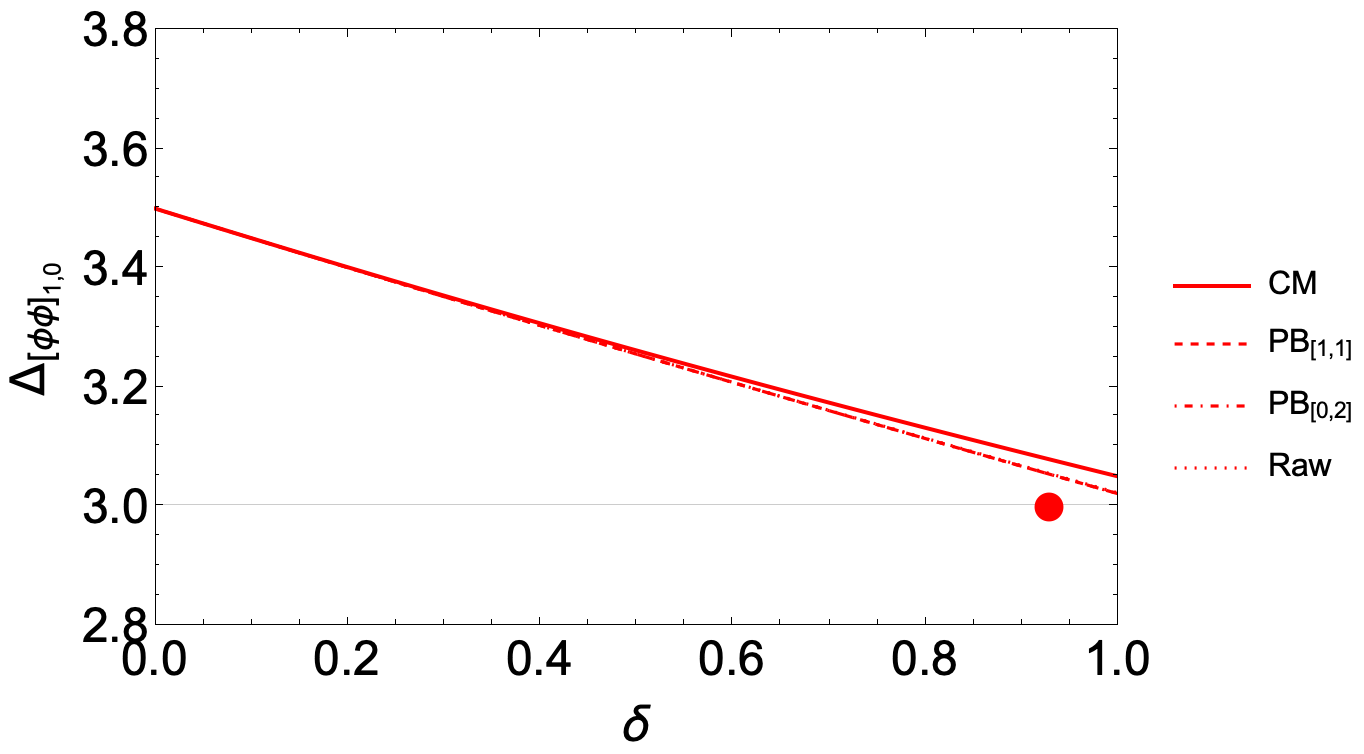}}
  
  \resizebox{0.6\columnwidth}{!}{\includegraphics{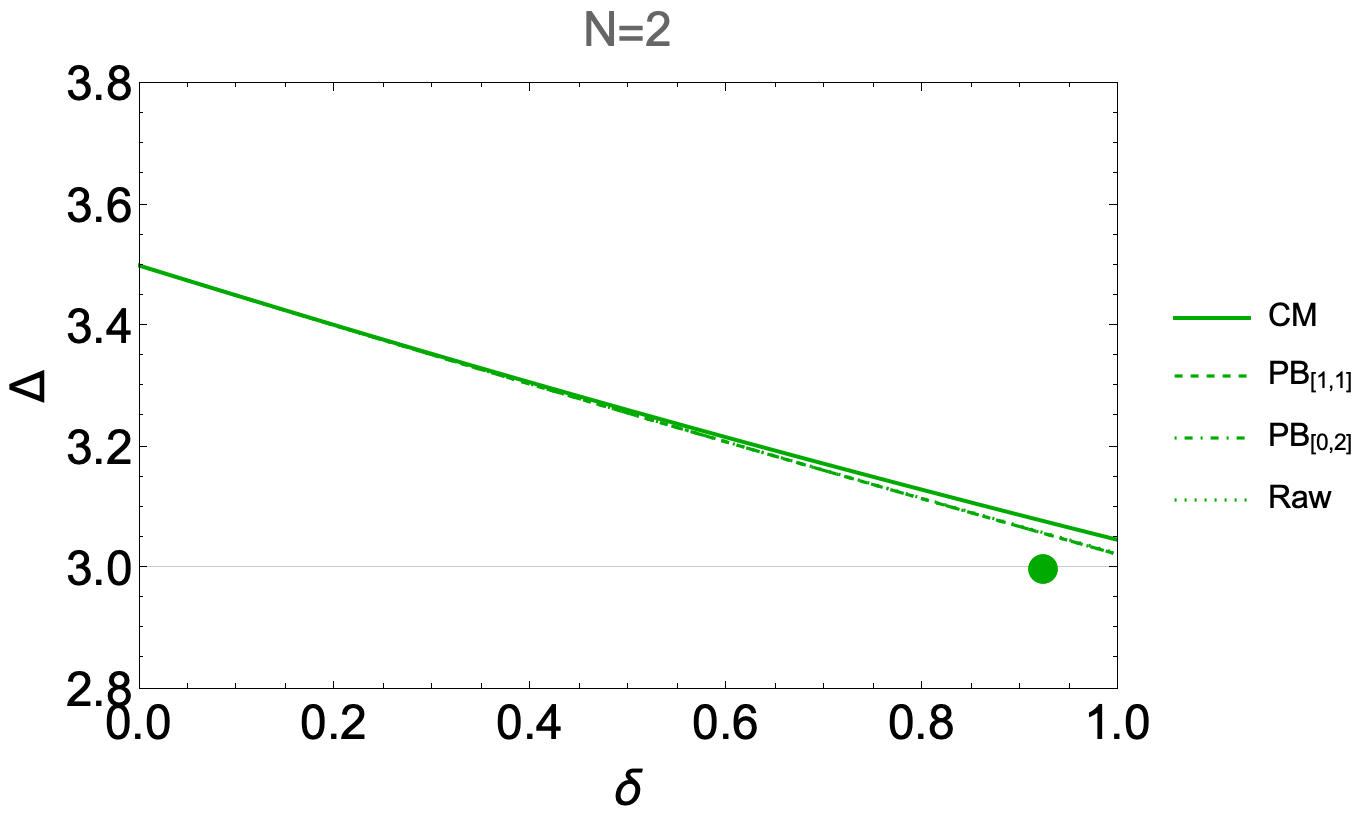}}
  
  \resizebox{0.6\columnwidth}{!}{\includegraphics{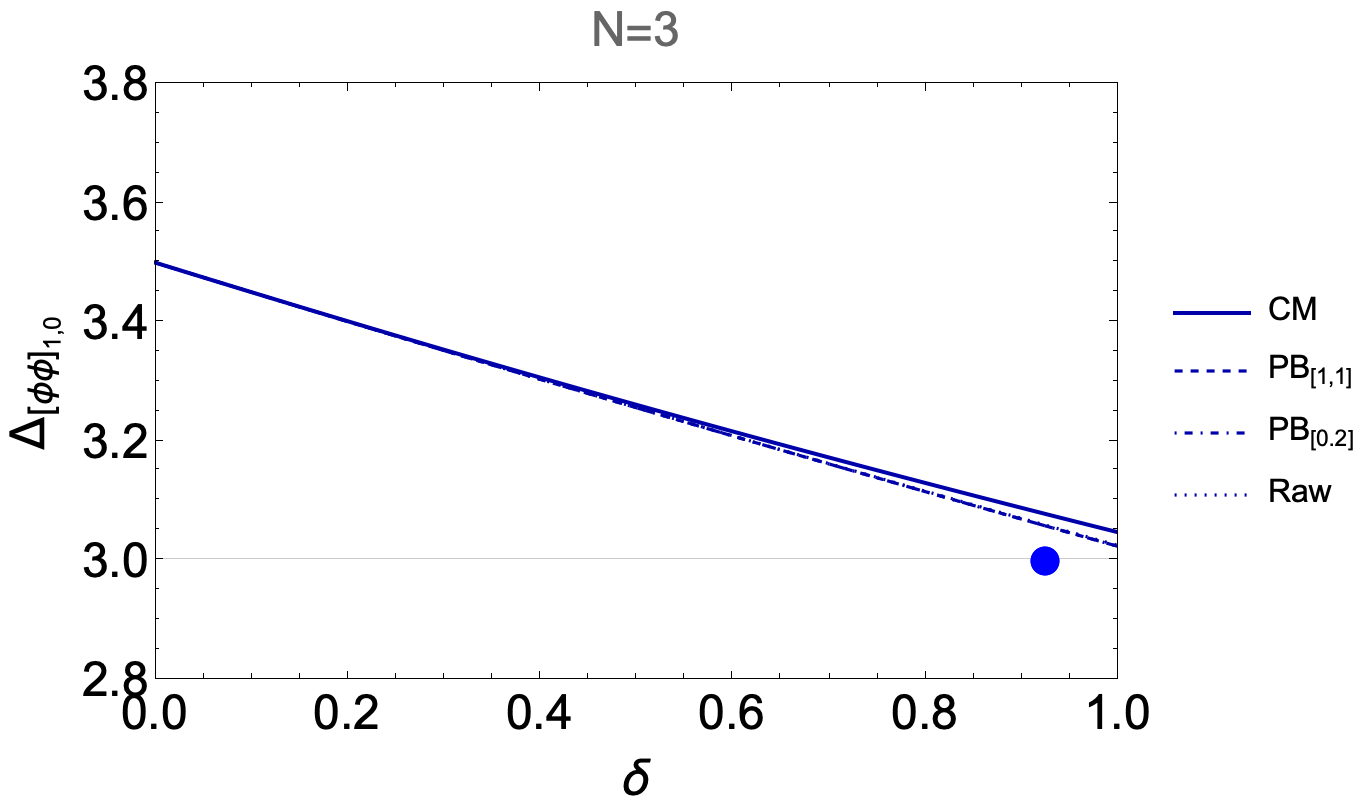}}
  
\end{center}
  \caption{Re-summation of the two-loop $\Delta_{[\phi\phi]_{1,0}}$ for $N = 1, 2, 3$. The solid line is the
  result of the re-summation using conformal mapping technique. The dashed and dash-dotted
  lines are the results using Pade$_{[1, 1]}$ and Pade$_{[0, 2]}$-Borel method. \ The dotted line is
  the result without re-summation. The dots indicate the results from conformal bootstrap \cite{kos2016precision,chester2020carving,chester2021bootstrapping}.\label{omegaplot2} }
\end{figure}

The re-summations of $\Delta_T$ and $\Delta_J$ are given in Fig \ref{deltaTJ}.
The re-summation of $\nu^{- 1} = 3 - \Delta_{\phi^2}$ is given in Fig
\ref{nuplot}. The re-summation of the critical exponents $\omega$ is given in
Fig. \ref{omegaplot}. 
The re-summation of the scaling dimension of $[\phi\phi]_{1,0}$ is given in
Fig. \ref{omegaplot2}. 
By comparing with the high precision results from conformal bootstrap \cite{kos2016precision,chester2020carving,chester2021bootstrapping}, we notice that even though we are
dealing with the perturbation series only in three-loop order, the re-summed
result is very good\footnote{We thank Ning Su for providing us the result $\omega=0.767$ for O(3) CFT.}. For comparison with the three-loop re-summation of the $4
- \epsilon$ result, see, for example, Chapter 17.4 of
{\cite{kleinert2001critical}}.

\

From Fig. \ref{deltaTJ}, we can read out $\delta$ for the short-range CFTs. We can then plug the corresponding $\delta$ into Fig
\ref{nuplot} to get the critical exponents $1/\nu$ for the short-range CFTs. We note down our prediction for the critical exponents in
Table \ref{exponents}.

\begin{table}[h]
\begin{center}
  \begin{tabular}{|l|l|l|l|}
    \hline
    & $\Delta_{\phi}$ & $\nu^{- 1}$ & $\omega_{\rm short}$\\
    \hline
    N=1 & $0.516 (8)$ & 1.59(1) & 0.78(22)\\
    \hline
    N=2 & $0.517 (8)$ & 1.51(2) & 0.79(22)\\
    \hline
    N=3 & 0.517(7) & 1.44(2) & 0.80(22)\\
    \hline
  \end{tabular}
\end{center}
  \caption{Critical exponents.\label{exponents}}
\end{table}

We now give some remarks about the critical exponents
$\omega_{\tmop{long}}$ for the long-range models. As we have explained in the previous section, at the long range limit $\delta \rightarrow 0$, we have two conformal primaries $\phi^4$ and $[\phi\phi]_{1,0}$. From Fig. \ref{omegaplot} and Fig. \ref{omegaplot2}, we see that their scaling dimension cross as we approach the short-range limit. The operator $\phi^4$ become the short range exponent $\omega_{\rm short}$, while $[\phi\phi]_{1,0}$ becomes marginal. The exponent $\omega_{\rm long}$ for the long-range model 
is given by the operator with lower dimension. Even though we observe level crossing in our re-summation, we expect that the scaling dimensions of the two operators to avoid crossing and show level-repulsion instead \cite{Korchemsky:2015cyx,Henriksson:2022gpa}.

The fact that the operator $[\phi\phi]_{1,0}$ becomes marginal is in accord with the recent study of long-range/short-range crossover in {\cite{Behan:2017emf,Behan:2017dwr}}. 
As already noticed in their works, the existence of a marginal operator will lead to logarithmic
finite size effect in Monte Carlo simulation, and therefore making the result converge
slowly. This explains the poor quality of the numerical results near the
crossover {\cite{Zhao:2023inj}}.

\section{Models with fermions}

\subsection{Models with four-fermion interactions in $D = 2 +
1$}\label{sec:3dfermion}

We now turn to fermionic models in D=3. We formulate our models using Majorana
fermions,
\begin{equation}
  \mathcal{L} = \mathcal{} \frac{1}{2} \mathcal{N}^{- 1} \bar{\psi}_a (-
  \partial)^{s - 2} \gamma^{\mu} \partial_{\mu} \psi_a + \lambda_Y
  (\overline{\psi_a} \psi_a)^2 + \lambda_T (\overline{\psi_a} \gamma^{\mu}
  \psi_b \Omega_{a b})^2 \label{3dfermionaction}
\end{equation}
Here we use the notation such that $\bar{\psi}_{\alpha} = \epsilon_{\alpha
\beta} \psi^{\beta}$, which is the Majorana conjugate available in $D = 2 + 1$
dimensions. We take the signature to be $\eta^{\mu \nu} = (- 1, 1, 1) .$ The
gamma matrices are
\[ \gamma^0 = i \sigma^2, \quad \gamma^1 = \sigma^1, \quad \gamma^2 =
   \sigma^3, \]
notice they are real matrices. The matrix $\epsilon . \gamma^{\mu}$ is
symmetric in our convention. The antisymmetric matrix
\[ \Omega =\mathbf{l}_{\frac{N}{2} \times \frac{N}{2} } \otimes \epsilon =
   \left( \begin{array}{ccccc}
     0 & 1 &  &  & \\
     - 1 & 0 &  &  & \\
     &  & \ldots . &  & \\
     &  &  & 0 & 1\\
     &  &  & - 1 & 0
   \end{array} \right), \]
is block diagonal. The kinetic term and the Gross-Neveu coupling term $g_Y
(\overline{\psi_i} \psi_i)^2$ preserves the O(N) symmetry. The Thirring
coupling $g_T (\overline{\psi_i} \gamma^{\mu} \psi_j \Omega_{i j})^2$ break
the symmetry to U(${\frac{N}{2}}$). One can re-write the action
\eqref{3dfermionaction} in a more familiar form using the Dirac fermions. The
Lagrangian \eqref{3dfermionaction} is the most general Lagrangian that
preserves the U(${\frac{N}{2}}$) symmetry. When $N = 1$, we have only two
grassmann variables, so that both the Gross-Neveu and the Thirring coupling
vanish, the theory is just the free theory. When $N = 2$, we have four
grassmann variables, the Gross-Neveu and the Thirring coupling are equivalent,
and we have only one coupling constant, which is a long-range generalization
of the original Thirring model {\cite{thirring1958soluble}}.

The short-range version of the above model was previously studied using the
functional renormalization group method in
{\cite{gies2010uv,janssen2012critical}}, using $2 + \epsilon$ expansion in
{\cite{Bennett:1999he,gracey1991computation,gracey1990three,Hikami:1976at}}.
The lattice version of the Thrring model was recently studied using Monte
Carlo simulation in
{\cite{del1996monte,del1999three,del1997three,hands1999phase,christofi2007critical,hands2019critical,Hands:2020itv,hands2023planar,hands2023spectroscopy}}.
The Thirring model is renormalizable in the large N limit {\cite{hands19951}}.
\ It has a hidden gauge symmetry {\cite{Itoh:1994cr}}, which we now explain.
The theory has a fixed point at which the Gross-Neveu coupling vanishes at the
large N limit, we get
\[ \mathcal{L} = \mathcal{} \frac{1}{2} i \bar{\psi}_a \gamma^{\mu}
   \partial_{\mu} \psi_a + \frac{\lambda}{N} (\overline{\psi_i} \gamma^{\mu}
   \psi_j \Omega_{i j})^2, \]
introduce the Hubbard-Stratonovich field $A_{\mu}$, we can write the above
action as
\begin{equation}
  \mathcal{L} = i \bar{\psi} \gamma \cdot \partial \psi - \sqrt{\frac{1}{N}}
  (\overline{\psi_a} \gamma^{\mu} \psi_a \Omega_{a b}) A'_{\mu} + \frac{1}{2
  \lambda} {A'}^2 . \label{thirringHS}
\end{equation}
One can view the field $A'_{\mu}$ as the massive vector boson from Stuckelberg
formalism. Just like the Gross-Neveu-Yukawa model is the UV completion of the
four-fermion Gross-Neveu model, the Thirring model has a UV completion given
by fermions coupled with a ``massive'' gauge field.

The long-range version of the Gross-Neveu model was first introduced in
{\cite{Chai:2021wac}}, and a large N study was performed in $1 < D < 4$ and
compared with the long-range Gross-Neveu-Yukawa model in $D = 4 - \epsilon$.
We will instead focus on the $s \rightarrow 2$ limit at $D = 3$, aiming to
explain how to recover short-range data from long-range perturbation theory.

The Feynman rules of the long-range model are the following. For each vertex,
we have
\[ V_{\alpha \beta \gamma \delta ; a b c d} = i \lambda_Y V^Y_{\alpha \beta
   \gamma \delta ; a b c d} + i \lambda_T V^T_{\alpha \beta \gamma \delta ; a
   b c d} \]
with
\[ V^T_{\alpha \beta \gamma \delta ; a b c d} = \Omega_{\tmop{ab}}
   \Omega_{\tmop{cd}}  (\epsilon . \gamma^{\mu})_{\alpha \beta} (\epsilon .
   \gamma_{\mu})_{\gamma \delta} + \tmop{permutations} \]
and
\[ V^Y_{\alpha \beta \gamma \delta ; a b c d} = \delta_{\tmop{ab}}
   \delta_{\tmop{cd}} \epsilon_{\alpha \beta} \epsilon_{\gamma \delta} +
   \tmop{permutations} \]
For each internal momentum, we need to integrate with
\[ \int \frac{d p^D}{(2 \pi)^D}, \]
For non-local theory, the propagator is given by
\[ G^{\alpha \beta}_{a b} (p) = \frac{- i p_{\mu} (\gamma^{\mu}
   \epsilon)^{\alpha \beta} + \mu \epsilon^{\alpha \beta}}{(p^2 + \mu^2)^{s /
   2}} \delta_{a b} . \]
We have introduced the IR regulator $\mu$ to make the Feynman integrals IR
finite. The final results will not depend on the mass. We take $s = \frac{D +
2 + \delta}{2}$, so that the scaling dimension of the scalar operator is
\begin{equation}
  \Delta_{\psi} = \frac{D - \delta}{4} . \label{fermiondimension}
\end{equation}
The four-fermion interaction is slightly irrelevant in the $\delta \rightarrow
0_-$ in the limit. Notice in this limit, the theory a (strongly) relevant
operator
\[ \bar{\psi}_a \gamma^{\mu} \partial_{\mu} \psi_a, \]
which is the short-range kinetic term for the fermions. We assume that we have
tuned the coupling of this term to zero. In other words, the fixed points that
we will discuss later are tri-critical points in the $\delta \rightarrow 0_-$
limit. The short-range models are located in the $\delta \approx - 1$ region, where above operator should become marginal, similar to the bosonic O(N) theory.

\

After evaluating all the Feynman diagram (given in Fig. \ref{betafermionic}
and Fig \ref{betafermionic2loop}) which contribute to the beta function
renormalization up to two loops, we get
\begin{eqnarray}
  \beta_Y & = & - \delta g_Y + \frac{1}{6 \pi^2} (g_Y^2 (4 \delta - 6 \delta
  \log (2) + N (\delta (\log (8) - 3) + 3) - 6) \nonumber\\&&- 2 (\delta (\log (64) - 5) +
  6) g_T^2 + 6 (\delta (\log (8) - 3) + 3) g_T g_Y) \nonumber\\&&+ \frac{1}{108 \pi^4}
  \bigg( 4 g_T^3 (- 205 \delta + 165 \delta \log (2) - 33 \delta \log (2) \log
  (64) + 41 \delta \log (64) \nonumber\\&&+ N (- \delta (\log (8) - 7) (\log (64) - 5) + 9
  \pi + 177 + 36 \log (2)) + 99 \pi + 489 + 396 \log (2)) \nonumber\\&&+ 3 g_T^2 g_Y  (-
  250 \delta - 378 \delta \log^2 (2) + 768 \delta \log (2)  \delta \log
  (4) - 24 \delta \log (64) + 24 \delta \log (4) \log (64)\nonumber\\&& + 3 N (\delta (- 6
  + \log (4) \log (8) - \log (16) \log (64) + \log (16777216)) + 9 \pi + 6 +
  90 \log (2) \nonumber\\&&  - 12 \log (4) - 2 \log (8) - 4 \log (64)) + 45 \pi + 354 - 396
  \log (2) + 144 \log (4) + 48 \log (64)) \nonumber\\&&+ 54 g_T g_Y^2  \big( \delta \big(
  12 N \log^2 (2) - N \log (4) \log (64) + 4 - 30 \log^2 (2) + 6 \log^2 (4) +
  \log (4) \big) \nonumber\\&& + 3 \pi + 4 + \log (4096) \big) + g_Y^3  (- 142 \delta +
  2 \delta \log^2 (8) + 44 \delta \log (8) \nonumber\\&&+ 9 N (\delta (2 - 6 \log^2 (2) +
  \log (16)) + 3 \pi + 6 + \log (4096)) - 9 \pi + 246 - 36 \log (2)) \bigg)
  \nonumber\\
  \beta_T & = & - \delta g_T + \frac{g_T  ((N + 2) (\delta (\log (8) - 7) + 3)
  g_T + 6 (\delta + \delta (- \log (8)) - 3) g_Y)}{18 \pi^2} \nonumber\\&&+ \frac{g_T}{324
  \pi^4} (g_T^2  (- 1958 \delta + 1192 \delta \log (8) - 510 \delta \log (2)
  \log (8) + 2 N (\delta (\log (8) - 7)^2 + 519 \nonumber\\&& - 18 \log (2)) - 9 \pi N + 765
  \pi + 4110 + 3060 \log (2)) + 6 g_T g_Y (- 194 \delta - 78 \delta \log (2)
  \log (8) \nonumber\\&&+ 148 \delta \log (8) + 2 N (- \delta (\log (8) - 7) (\log (64) -
  5) + 9 \pi + 141 + 9 \log (16)) + 117 \pi + 834 \nonumber\\&&+ 468 \log (2)) + 9 g_Y^2 
  (- 118 \delta - 18 \delta \log^2 (2) + 144 \delta \log (2) + 3 N (\delta (-
  2 - 6 \log^2 (2)  \nonumber\\&& + \log (256)) + 3 \pi + 10 + \log (4096)) + 9 \pi + 270 + 6
  \log (64))) . \label{3dfermionbeta} 
\end{eqnarray}
Here $g_Y$ and $g_T$ are the renormalized couplings of the bare couplings
$\lambda_Y$ and $\lambda_T$, for there definition, see
\eqref{3Dfermioncoupling}. The scaling dimension of the stress tensor operator
is given by
\begin{equation}
  \Delta_{T^{\mu \nu}} = 2 \Delta_{\psi} + 1 - \frac{\Gamma \left( \frac{1}{4}
  \right)  (3 Ng_T^2 + Ng_Y^2 + 6 g_T g_Y + 3 g_T^2 - g_Y^2)}{60 \pi^4 \Gamma
  \left( \frac{5}{4} \right)} . \label{3dfermionT}
\end{equation}
The scaling dimension of the fermion mass operator $\overline{\psi_i} \psi_i$
is
\begin{eqnarray}
  \Delta_{\bar{\psi} \psi} =&& \frac{3 - \delta}{2} + \frac{(\delta (\log (64) -
  6) + 6) g_T}{4 \pi^2} + \frac{g_Y  ((N - 1) (\delta (\log (4) - 2) + 2))}{4
  \pi^2} \nonumber\\&&+ \frac{g_T g_Y}{12 \pi^4} (58 \delta - 2 \delta \log (8) (4 + \log
  (8)) + 16 N (\delta (\log (8) - 4) + 3) + 9 \pi - 114 + 36 \log (2))\nonumber\\&& +
  \frac{(N - 1) g_Y^2 }{72 \pi^4} (58 \delta - 2 \delta \log (8) (4 + \log
  (8)) + 16 N (\delta (\log (8) - 4) + 3) + 9 \pi \nonumber\\&&- 114 + 36 \log (2)) +
  \frac{g_T^2 }{8 \pi^4} (- 2 \delta (33 + \log (2) (\log (8) - 28)) + N (- 2
  \delta (\log (2) - 1) (\log (8) - 1) \nonumber\\&&+ 3 \pi - 22 + \log (4096)) + 3 \pi +
  26 + \log (4096)) . \label{3dfermionMass}
\end{eqnarray}
The Feynman integrals we encounter are given in Appendix
\ref{thirringdetails}. One can check that there is no wavefunction
renormalization since the kinetic term is non-local. The fixed points of the
one-loop beta functions are
\begin{eqnarray}
  F_1 & : & g_Y = g_T = 0, \nonumber\\
  F_2 & : & g_Y = \frac{2 \pi^2}{N - 2} \delta, \quad g_T = 0, 
  \label{GNfixedpoint}\\
  F_3 & : & g_Y = - \frac{\pi^2 }{N^3 + 2 N^2 + 32 N - 80} \left( - N^2 + (N +
  2) \sqrt{N^2 + 112 N + 256} + 14 N - 112 \right) \delta,\nonumber\\&&  g_T = \frac{6
  \pi^2 }{N^3 + 2 N^2 + 32 N - 80} \left( N^2 - \sqrt{N^2 + 112 N + 256} + N +
  16 \right) \delta \nonumber\\
  F_4 & : & g_Y = \frac{\pi^2 }{N^3 + 2 N^2 + 32 N - 80} \left( N^2 + (N + 2)
  \sqrt{N^2 + 112 N + 256} - 14 N + 112 \right) \delta, \nonumber\\&&  g_T = \frac{6
  \pi^2 }{N^3 + 2 N^2 + 32 N - 80} \left( N^2 + \sqrt{N^2 + 112 N + 256} + N +
  16 \right) \delta . \nonumber
\end{eqnarray}
The free fixed point $F_1$ is the most stable fixed point without relevant
direction. The second fixed point $F_2$ is the long-range generalization of
the Gross-Neveu model, for large enough $N$, the stability matrix
\[ \frac{\partial \beta_i}{\partial g_j} \]
has one positive eigenvalue, which means that the RG flow has one relevant
direction. The third fixed point $F_3$ is the long-range generalization of the
Thirring model, for large enough $N$, the fixed point has one relevant RG
directions. The fourth fixed point has two irrelevant direction, which is a
tri-critical point. The fixed point structure resembles those found in
{\cite{gies2010uv,janssen2012critical}} for the short-range model.

Notice that the coupling $g_Y$ of the Gross-Neveu fixed point diverges at $N =
2$, this is analogous to
{\cite{wetzel1984two,gracey1991computation,gracey1990three}}. This is because
the Thirring coupling is equivalent to the Gross-Neveu coupling, as we have
mentioned. The theory at $N = 2$ needs special treatment, we can set $g_T =
0$, since the two couplings are equivalent, the beta function then becomes
\begin{eqnarray}
 \beta_Y =&& - \delta g_Y - \frac{\delta g_Y^2}{3 \pi^2} + \frac{g_Y^3}{108
   \pi^4} (2 \delta (- 53 - 54 \log^2 (2) + \log^2 (8) + 22 \log (8)\nonumber\\&& + 9 \log
   (16)) + 45 \pi + 354 - 36 \log (2) + 18 \log (4096)) . 
\end{eqnarray}
 The critical point becomes
\[ g_{\star} = \pm 4.11877 \sqrt{\delta} + \mathcal{O} (\delta^1) . \]
To solve the fixed point to the next order, we need the beta function at
$g_Y^4$ order (three-loop), which we leave for future work. The scaling
dimensions are
$$ \Delta_T = 2.5 - (0.5 \pm 0.01161) \delta + \mathcal{O} (\delta^{3 / 2}),
   \quad {\rm and} \quad {{\Delta_{\bar{\psi} \psi}}  =
   \frac{3}{2}} \pm 0.208659 \sqrt{\delta} + \mathcal{O} (\delta^1) . $$
At small $\delta$, these fixed points are clearly not smoothly connected to
the fixed point \eqref{GNfixedpoint} as we vary $N$. For large $\delta$, it
remains an open question whether one can find a short-range CFT, which is
smoothly connected to the short-range CFTs with large $N$.

\

We now discuss the $N > 2$ solutions. Now the question is whether the fixed
points will lead to unitary CFTs. It is too early to try to infer the final
result just from a naive two-loop calculation, we will however plot the result
just to get a feeling of it. \ For the Gross-Neveu fixed point, we have
\begin{eqnarray*}
  \text{Gross-Neveu} : & N = 4 & \Delta_T = 2.5 - 0.5 \delta - 0.2 \delta^2,
  \hspace{3em} {\Delta_{\bar{\psi} \psi}}  = 1.5 + \delta -
  8.58085 \delta^2,\\
  & N = 8 & \Delta_T = 2.5 - 0.5 \delta - 0.0518519 \delta^2,
  {\Delta_{\bar{\psi} \psi}}  = 1.5 + 0.666667 \delta + 1.33808 \delta^2 .
\end{eqnarray*}

\begin{figure}[h]
  \resizebox{0.5\columnwidth}{!}{\includegraphics{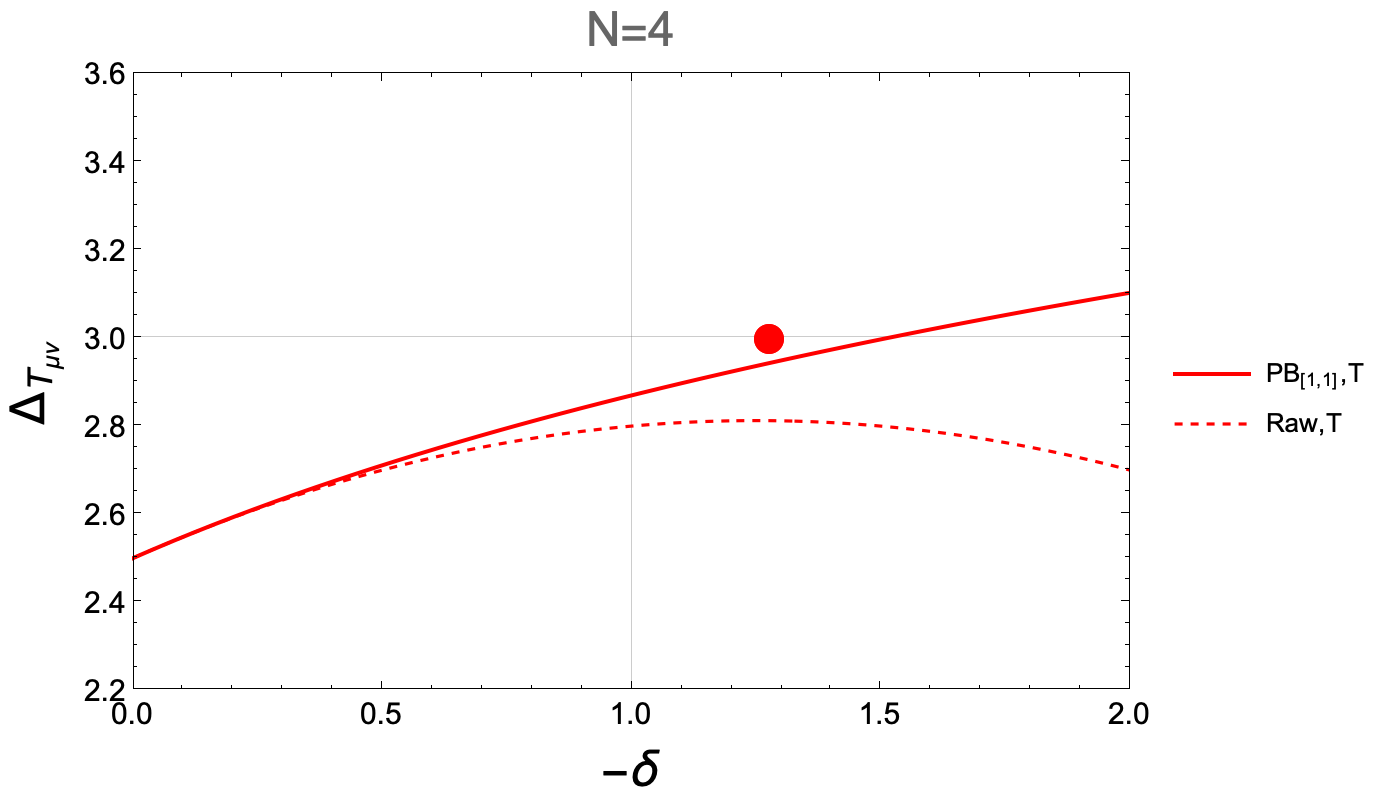}}\resizebox{0.5\columnwidth}{!}{\includegraphics{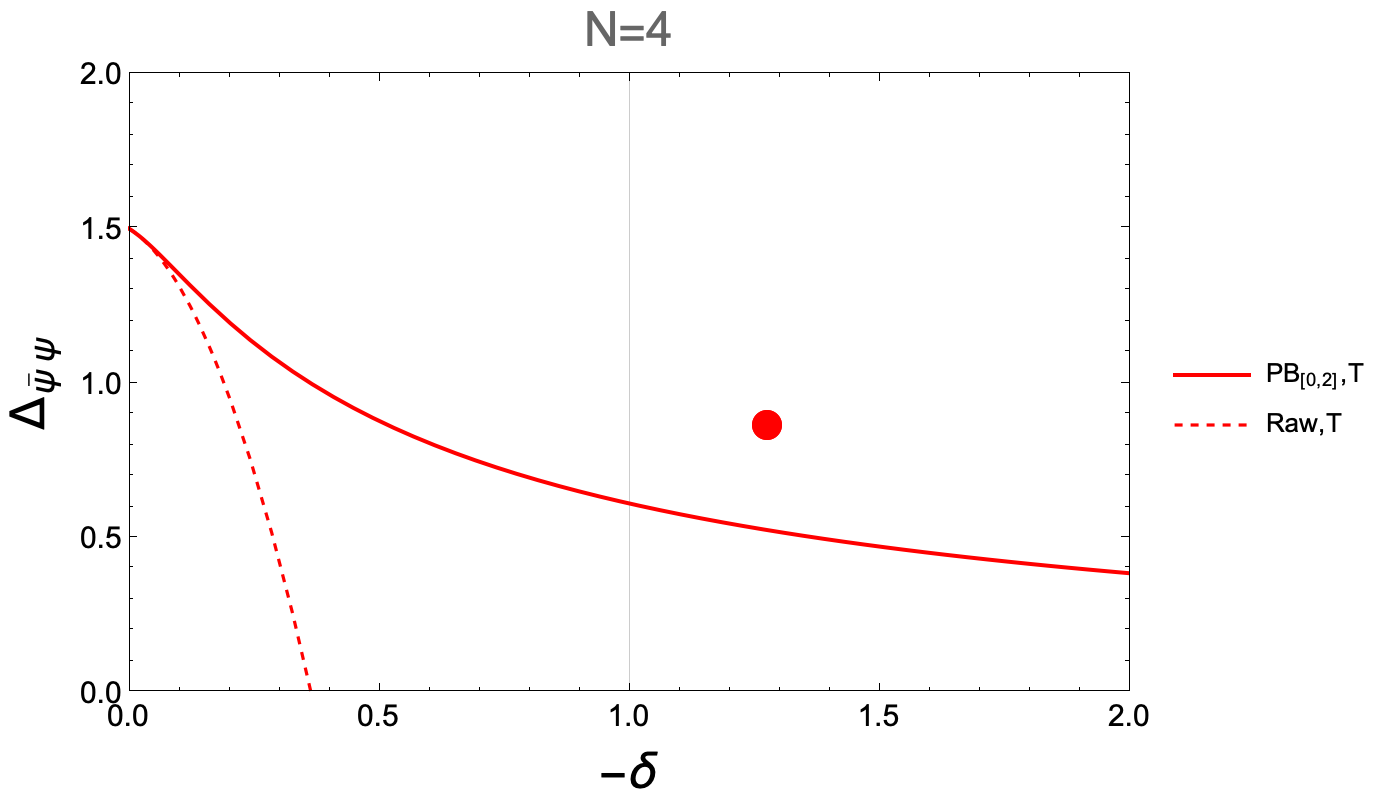}}
  
  \resizebox{0.5\columnwidth}{!}{\includegraphics{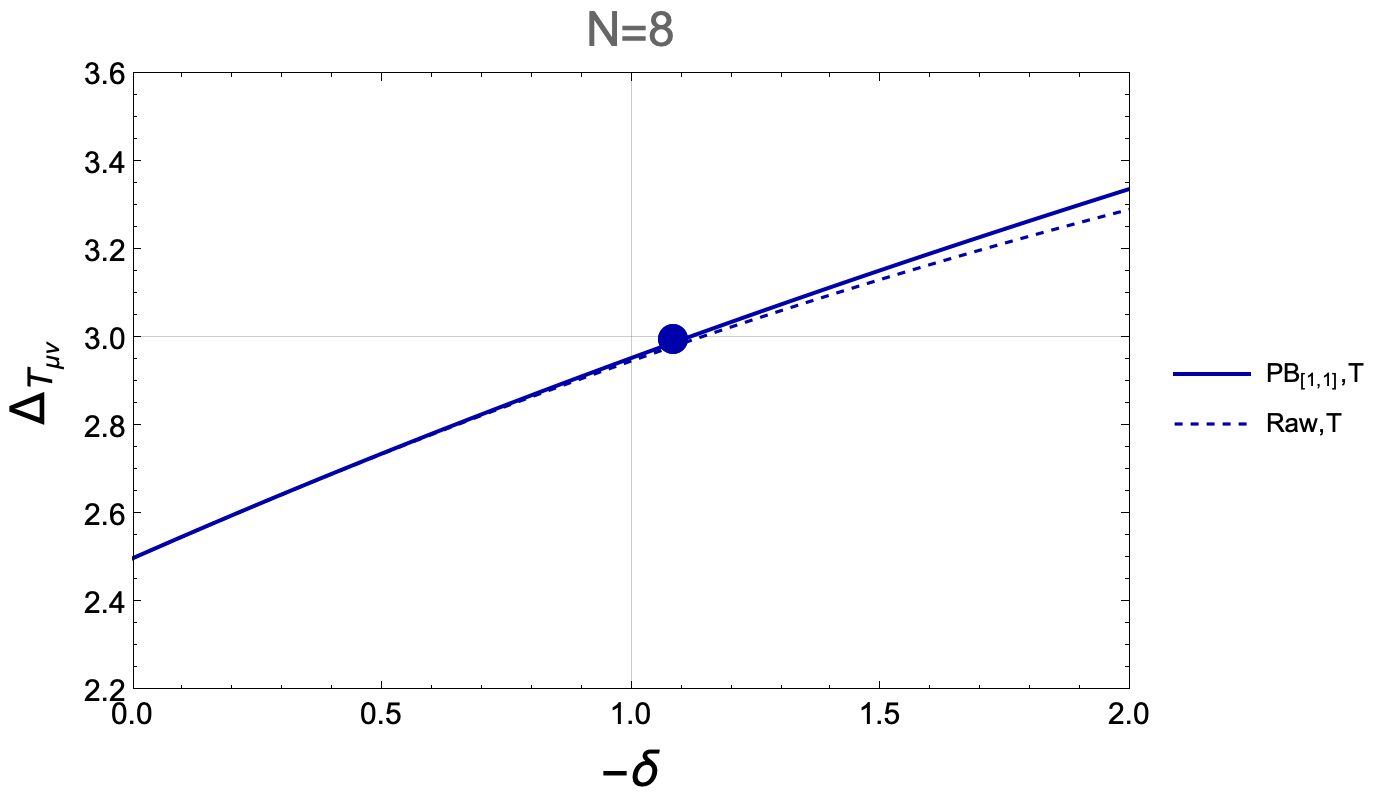}}\resizebox{0.5\columnwidth}{!}{\includegraphics{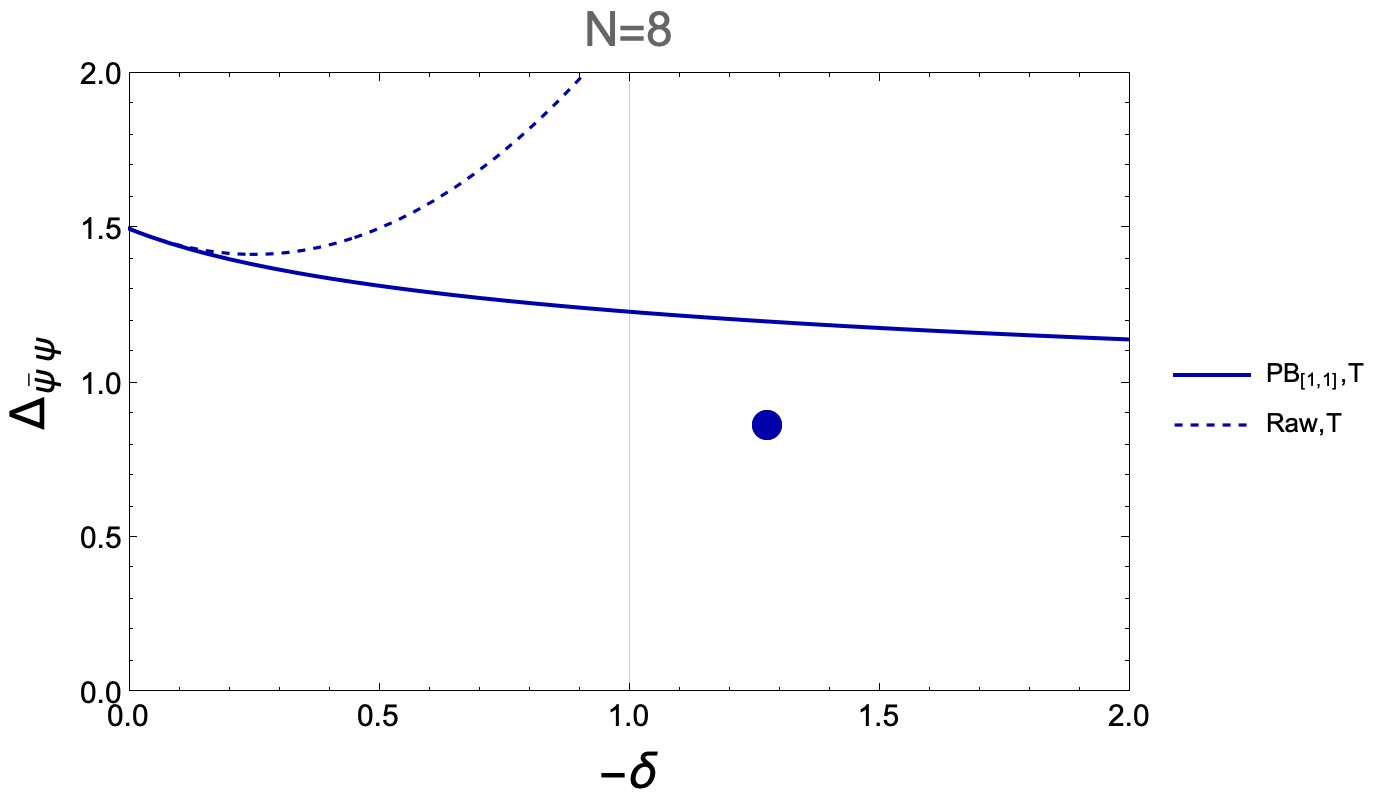}}
  \caption{$\Delta_{T_{\mu \nu}}$ and $\Delta_{\bar{\psi} \psi}$ of the
  Gross-Neveu model. We have used Pade-Borel re-summation method. Sometimes,
  the Pade approximations of the corresponding Borel functions have poles with
  positive real parts, we discard these cases. The dots indicate the results from conformal bootstrap \cite{erramilli2023gross}. \label{GN3D}}
\end{figure}

The Gross-Neveu(-Yukawa) model has been solved by the conformal bootstrap
method. For the Thirring model, on the other hand, the long-range perturbation
theory may allow us to be ahead of the bootstrap method. For the Thirring
fixed point, we have,
\begin{eqnarray*}
  \tmop{Thirring} : & N = 8 & \Delta_T = 2.5 - 0.5 \delta - 1.7714 \delta^2,
  \hspace{2em} {\Delta_{\bar{\psi} \psi}}  = 1.5 + 2.62541 \delta - 326.947
  \delta^2,\\
  & N = 100 & \Delta_T = 2.5 - 0.5 \delta - 0.0772493 \delta^2, \hspace{1em}
  {\Delta_{\bar{\psi} \psi}}  = 1.5 + 0.734632 \delta - 4.81543 \delta^2 .
\end{eqnarray*}
The re-summed curve $\Delta_T (\delta)$ will intersect with the $\Delta_T = 3$
line when $N  > 12$. However, when $N$ is close to 12, the intersection happens at $|\delta| \gg 1$ , we should not trust our perturbation theory in that region. When $N>29$, the intersection happens at $|\delta| \approx 1$ We therefore take $N_c\approx29$ as our two-loop prediction of the critical $N_c$ above which the short-range the Thirring fixed point exists.

\begin{figure}[h]
  \resizebox{0.5\columnwidth}{!}{\includegraphics{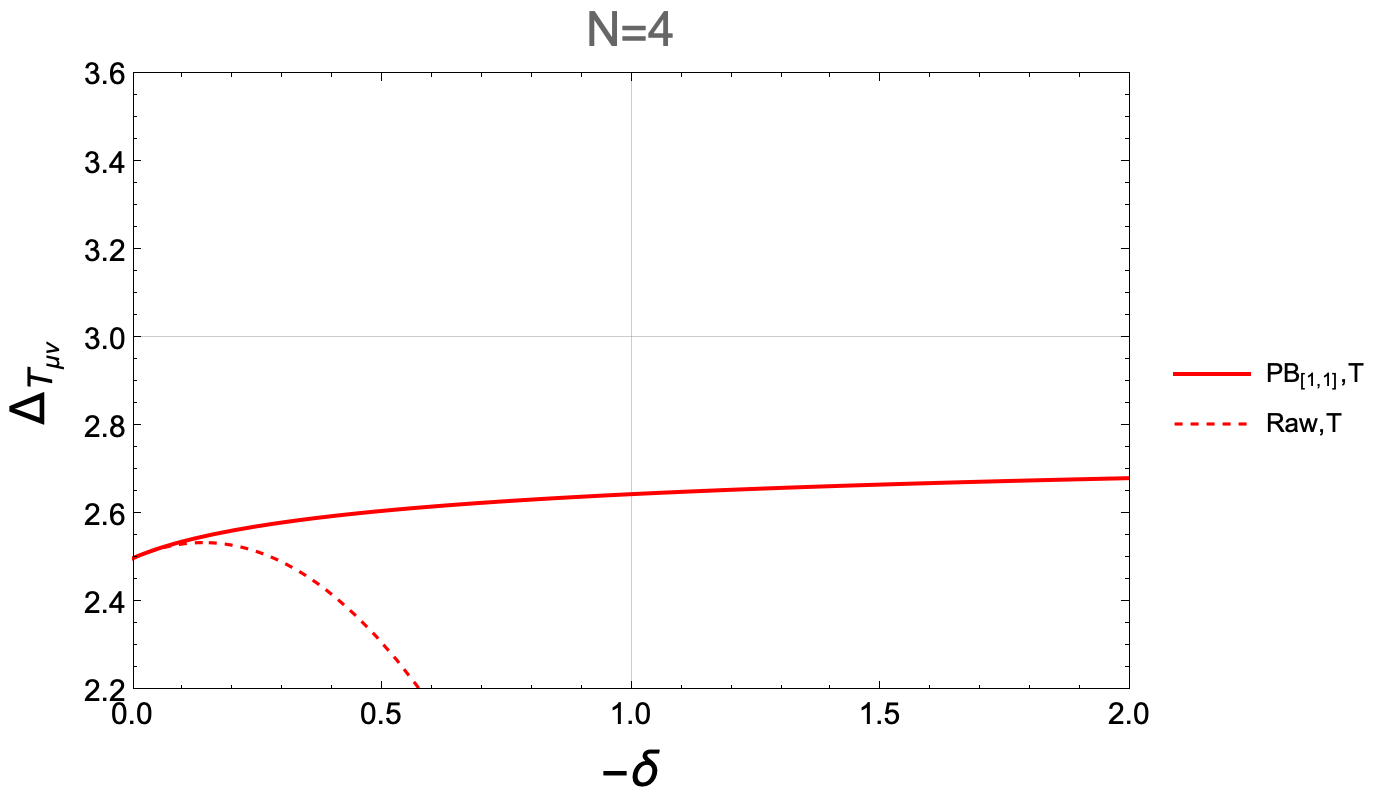}}\resizebox{0.5\columnwidth}{!}{\includegraphics{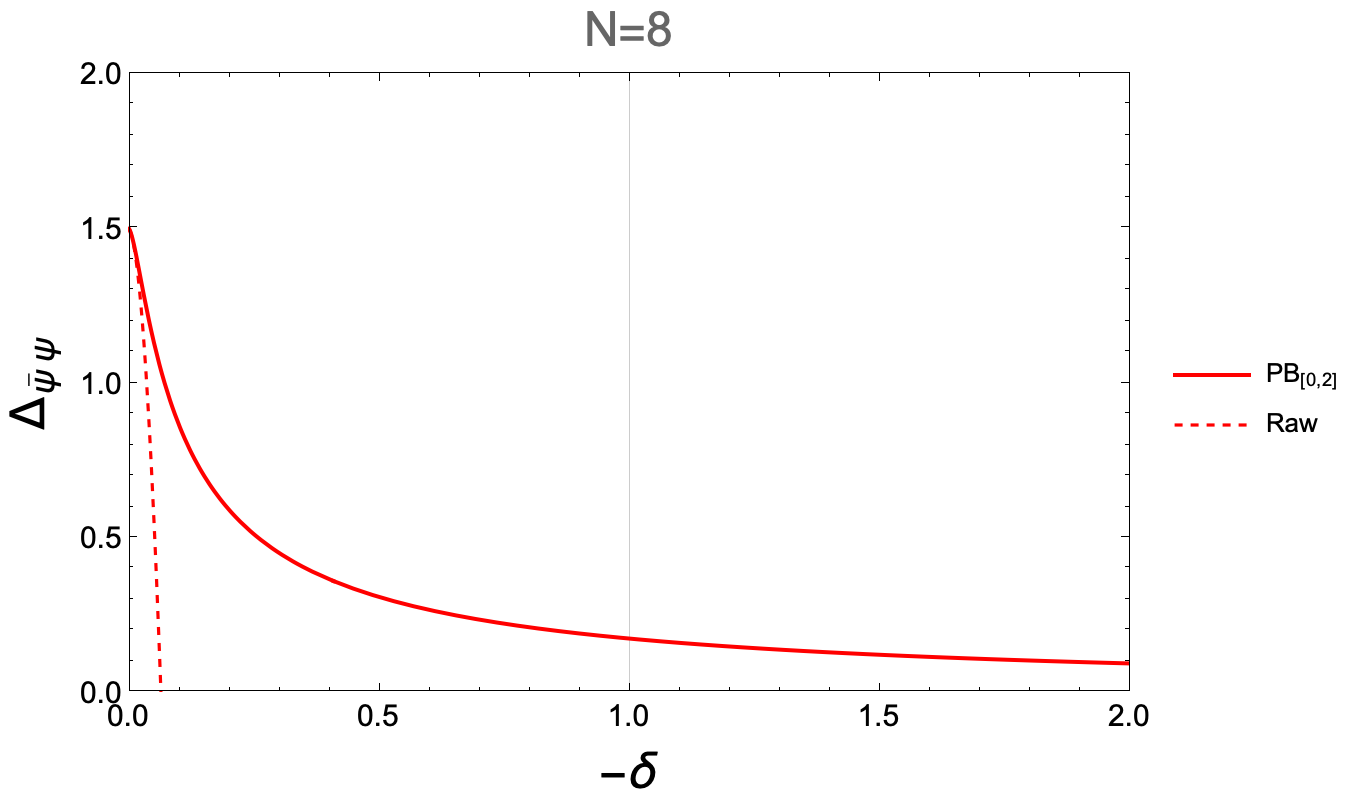}}
  
  \resizebox{0.5\columnwidth}{!}{\includegraphics{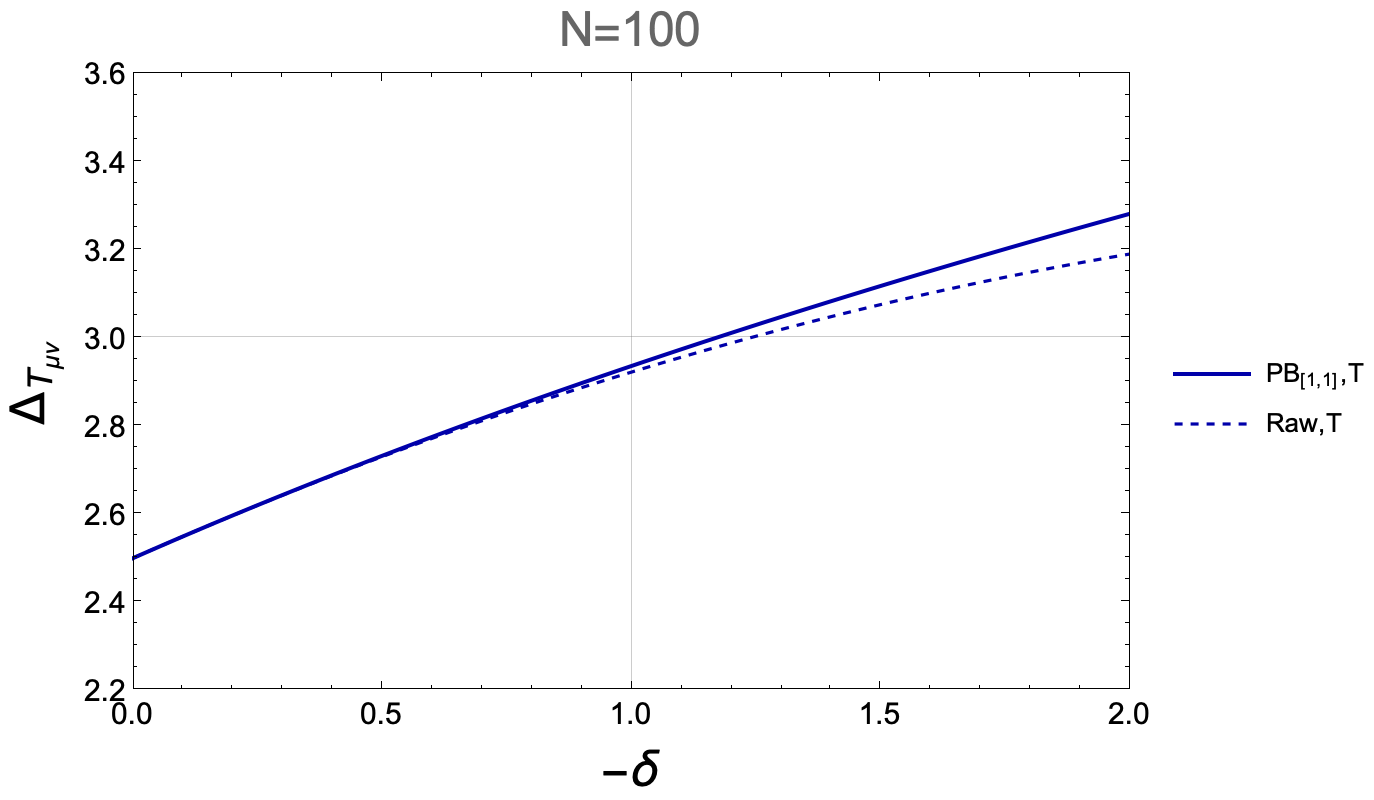}}\resizebox{0.5\columnwidth}{!}{\includegraphics{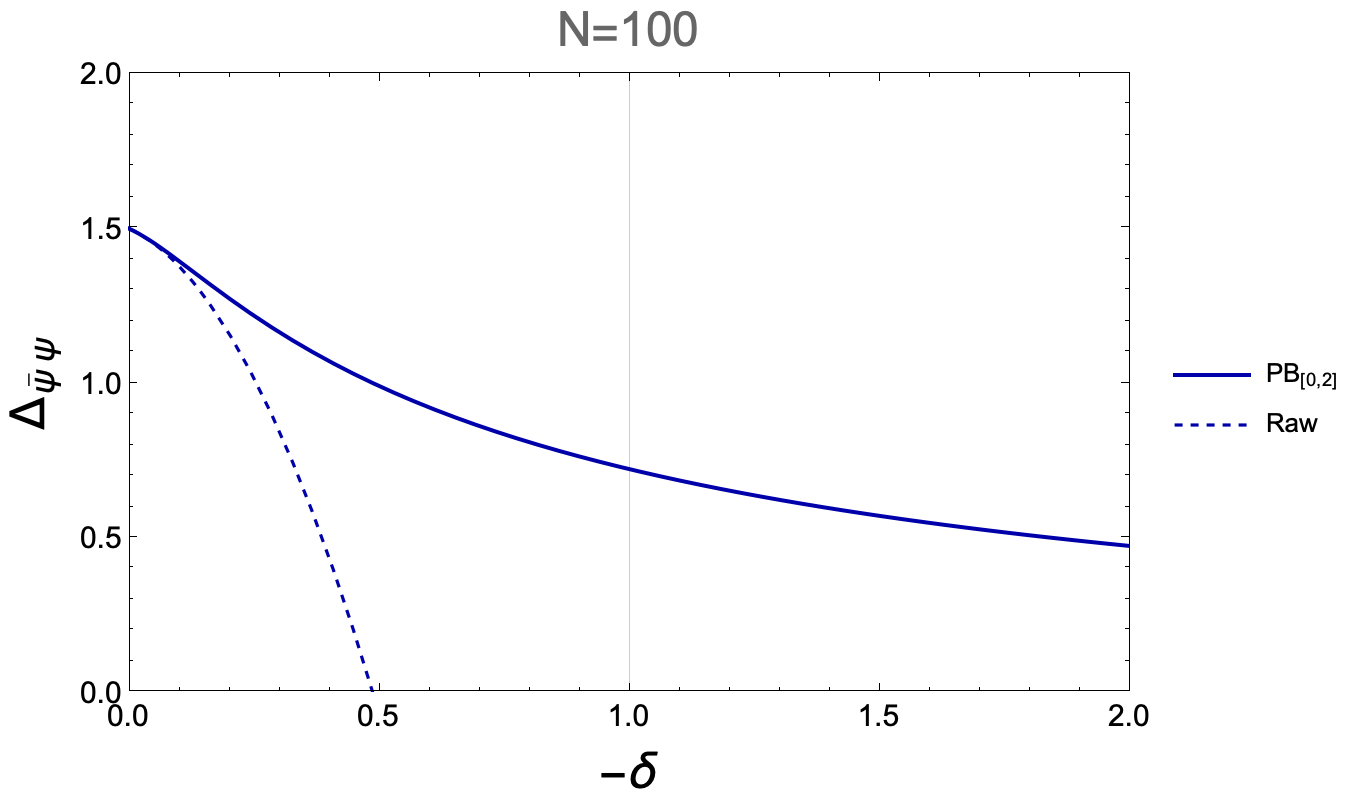}}
  \caption{The Thirring fixed point from $\Delta_T = 3$. We have used
  Pade-Borel re-summation method. Sometimes, the Pade approximations of the
  corresponding Borel functions has poles with positive real parts, we discard
  these cases.\label{TR3D}}
\end{figure}

Notice that at the $\delta\rightarrow0$ limit, $\Delta_{\psi}$ is below the unitarity bound. It is important to check that when $\Delta_T=D$ happens, the corresponding $\Delta_{\psi}$ is above the unitarity bound, so that the short-range model is unitary. One can see in Fig. \ref{GN3D} and \ref{TR3D} that this is indeed the case.

\subsection{Generalized Thirring model in $D = 4 + 1$}\label{sec:5dfermion}

We also consider a fermionic model in D=4+1. We formulate our theory using
symplectic Majorana fermions (a Dirac fermion can be written as two symplectic
Majorana fermions). A nice review of the Clifford algebras in general
dimensions is given in {\cite{freedman2012supergravity}}. We will study the
following model,
\begin{eqnarray}
  \mathcal{L} & = & \frac{1}{2} \mathcal{N}^{- 1} \psi_{i, \alpha} (C. 
  \gamma^{\mu})^{\alpha \beta} (- \partial)^{s - 2} \partial_{\mu} \psi_{j
  \beta} \Omega_{i j} \nonumber \\
  &  & + \frac{1}{8} \lambda_{T_2} (\psi_{i, \alpha} (C.  \gamma^{\mu
  \nu})^{\alpha \beta} \psi_{j, \beta} \Omega_{i j}) (\psi_{i, \alpha} (C. 
  \gamma_{\mu \nu})^{\alpha \beta} \psi_{j, \beta} \Omega_{i j}), \quad
  \tmop{with} \quad i = 1 \ldots N. \nonumber \\\label{5dThirring} 
\end{eqnarray}
Clearly, $N$ is an even number. We have also defined that $\gamma^{\mu \nu}
\equiv \gamma^{[\mu \nu]} = \gamma^{\mu} \gamma^{\nu} + \gamma^{\nu}
\gamma^{\mu}$ . We take the convention of gamma matrix to be
\begin{eqnarray*}
  &  & \gamma^0 = i \sigma^1 \otimes \mathbf{l}_{2 \times 2}, \quad \gamma^1
  = \sigma^2 \otimes \mathbf{l}_{2 \times 2},\\
  &  & \gamma^2 = \sigma^3 \otimes \sigma^1, \quad \gamma^3 = \sigma^3
  \otimes \sigma^2,\\
  &  & \gamma^4 = \sigma^3 \otimes \sigma^3, \quad B = -\mathbf{l}_{2 \times
  2} \otimes \sigma^2,\\
  &  & C = B^T . \gamma^0 .
\end{eqnarray*}
The symplectic Majorana fermions are defined as
\[ \psi_i = \Omega_{i j} B^{- 1} (\psi_j)^{\star} . \]
In our convention, the matrix $C$ and $(C.  \gamma^{\mu})$ are antisymmetric,
while $(C.  \gamma^{\mu \nu})$ is symmetric. The gamma matrices satisfy
\[ B^T . (C)^{\star} .B = - C, \quad B^T . (C.  \gamma^{\mu})^{\star} .B = C.
   \gamma^{\mu}, \quad B^T . (C.  \gamma^{\mu \nu})^{\star} .B = - C. 
   \gamma^{\mu \nu}, \]
so that the action is Hermitian. We can consider other four-fermion couplings
such as Gross-Neveu coupling $\frac{1}{8} \lambda_Y (\psi_{i, \alpha}
(C)^{\alpha \beta} \psi_{i, \beta})^2$, and $\frac{1}{8}
\lambda_T (\psi_{i, \alpha} (C.  \gamma^{\mu})^{\alpha \beta} \psi_{i, \beta})
(\psi_{i, \alpha} (C.  \gamma_{\mu})^{\alpha \beta} \psi_{i, \beta})$. These
terms, however, will break the $\tmop{Sp} (N)$ symmetry to the $U \left(
\frac{N}{2} \right)$ group. The above theory is the maximally symmetric action
in $D = 4 + 1$.

Similar to the $D = 2 + 1$ dimensional case, we take $s = \frac{D + 2 +
\delta}{2}$, so that the scaling dimension of the scalar operator is
\[ \Delta_{\psi} = \frac{D - \delta}{4} . \]
The four fermion interaction is slightly irrelevant in the $\delta \rightarrow
0_-$ in the limit. We assume that we have tuned the coupling of the
short-range kinetic term
\[ \psi_{i, \alpha} (C.  \gamma^{\mu})^{\alpha \beta} \partial_{\mu} \psi_{j
   \beta} \Omega_{i j} \]
to zero.
The Feynman rule for the propagators are
\[ G_{\alpha \beta, i j} (p) = - i \frac{p_{\mu} (\gamma^{\mu} C)_{\alpha
   \beta} \Omega_{i j} - \mu C_{\alpha \beta} \delta_{i j}}{(p^2 + \mu^2)^{s /
   2}} . \]
We have introduced the mass term to make the Feynman integrals IR finite,
which corresponds to add $- i \frac{1}{2} \mu \psi_i C \psi_i \delta_{i j}$ to
the action. The final results will not depend on the mass. The four-fermion
vertice is
\begin{eqnarray*}
  &  & V^{T_2}_{\alpha \beta \gamma \delta ; a b c d}\\
  & = & i \lambda_{T_2} (\Omega_{ad} \Omega_{bc} (C. \gamma^{\mu
  \nu})_{\alpha \delta} (C. \gamma_{\mu \nu})_{\beta \gamma} + \Omega_{ab}
  \Omega_{cd} (C. \gamma^{\mu \nu})_{\alpha \beta} (C. \gamma_{\mu
  \nu})_{\gamma \delta} + \Omega_{ac} \Omega_{bd} (C. \gamma^{\mu
  \nu})_{\alpha \gamma} (C. \gamma_{\mu \nu})_{\beta \delta}),
\end{eqnarray*}
The beta function at 2-loop order looks like,
\begin{eqnarray}
  \beta_{T_2} & = & - \delta g_{T_2} + \frac{g_{T_2}^2  \left( \frac{4}{15}
  \delta (4 N - 337) - \delta (N - 28) \log (4) - 2 N + 56 \right)}{15 \pi^3}
  \nonumber\\&&+ \frac{4 g_{T_2}^3 }{50625 \pi^6} (64296 \delta - 21952 \delta N - 450
  \delta (59 N + 3) \log^2 (2) \nonumber\\&&+ 60 \log (2) (- 816 \delta + (922 \delta +
  885) N + 45) + 23160 N - 13275 \pi N - 675 \pi - 2280) .\nonumber\\
  \label{5Dfermionbeta} 
\end{eqnarray}
Here $g_{T_2}$ is the renormalized coupling of $\lambda_{T_2}$. The anomalous
dimension of the stress tensor operator $T_{\mu \nu}$ is
\[ \Delta_{T_{\mu \nu}} = \frac{7}{2} - \frac{\delta}{2} - \frac{4 (N + 1)
   \Gamma \left( \frac{3}{4} \right) g_{T_2}^2}{21 \pi^6 \Gamma \left(
   \frac{7}{4} \right)} . \label{5DfermionT} \]
The anomalous dimension of the mass operator $\psi_{i, \alpha} (C)^{\alpha
\beta} \psi_{i, \beta}$ is
\begin{eqnarray}
  \Delta_{\bar{\psi} \psi} =&& \frac{5}{2} - \frac{\delta}{2} + \frac{5 (\delta
  (\log (64) - 8) + 6) g_{T_2}}{9 \pi^3} + \frac{2 g_{T_2}^2 }{405 \pi^6} (-
  4128 \delta - 64 \delta N - 90 \delta (N - 3) \log^2 (2) \nonumber\\&&+ 12 \log (2) (183
  \delta + (14 \delta + 15) N - 45) - 72 N - 45 \pi N + 135 \pi + 2916) .
  \label{5DfermionMass}
\end{eqnarray}
When $N \neq 28$, at the fixed point, we have
\[
\Delta_{T_{\mu \nu}} = \frac{7}{2} - \frac{\delta}{2} - \frac{75 \delta^2 
   (N + 1) \Gamma \left( \frac{3}{4} \right)}{7 (N - 28)^2 \Gamma \left(
   \frac{7}{4} \right)} . \]
\begin{eqnarray}
 \Delta_{\bar{\psi} \psi} = &&\frac{5}{2} + \frac{\delta (N + 22)}{56 - 2 N} +
   \frac{\delta^2 }{6 (N - 28)^3} (80 (268 N - 2119) \nonumber\\&&- 75 \pi (N (N + 87) +
   90) + 300 (N (N + 87) + 90) \log (2)) . 
\end{eqnarray}
When $N = 28$, the solution is at
\[ g_{\star} = \pm 4.89026 \sqrt{\delta} + \mathcal{O} (\delta^1) . \]
The scaling dimensions are
\[ \Delta_T = 3.5 - (0.5 \pm 0.183207) \delta + \mathcal{O} (\delta^{3 / 2}),
   \quad {\Delta_{\bar{\psi} \psi}}  = \frac{5}{2} \pm 0.525728 \sqrt{\delta}
   + \mathcal{O} (\delta) . \]
Just like the 2+1 dimensional Thirring model can be though of as the fermions
coupled Stuckelberg vector field {\cite{Itoh:1994cr}}, the 4+1 dimension model
\eqref{5dThirring} can be though as fermions coupled with ``Higgsed'' two-form
fields. In other words, one can introduce the auxiliary field $B_{\mu \nu}$
\begin{eqnarray}
  \mathcal{L} & = & \frac{1}{2} \psi_{i, \alpha} (C.  \gamma^{\mu})^{\alpha
  \beta} \partial_{\mu} \psi_{j \beta} \Omega_{i j} - \sqrt{\frac{1}{N}}
  (\psi_{i, \alpha} (C.  \gamma^{\mu \nu})^{\alpha \beta} \psi_{j, \beta}
  \Omega_{i j}) B_{\mu \nu} \nonumber\\
  &  & + \frac{4}{\lambda_{T_2}} (\psi_{i, \alpha} (C.  \gamma^{\mu
  \nu})^{\alpha \beta} \psi_{j, \beta} \Omega_{i j}) B_{\mu \nu} \quad
  \tmop{with} \quad i = 1 \ldots N. \label{5dThirringHS} 
\end{eqnarray}
and think about $B_{\mu \nu}$ as a dynamical ``massive'' two form field. It
will be interesting to study the above model in the large $N$ limit.

\section{Discussion}

The main point of our paper is that by imposing the conformal Ward identity,
we can extract conformal data of local/short-range CFTs from long-range
perturbation theory. We have applied this idea to the O(N) vector models. It
is interesting to think about how to apply this idea to other quantum field
theories.

\

1. {{the multi-scalar theories, irreducible fixed points.}} In the O(N)
vector model, all the scalar fields transform in a single irreducible
representation of the O(N) group. It would be interesting to consider smaller
symmetry groups so that the scalars transform in more than one irreducible
representations. In this case, scalars can have different scaling dimensions,
to take this into account, we consider the following theory
\[ \mathcal{L}= \int d^D xd^D y \sum_{a = 1}^N \frac{\phi^a (x) \phi^a (y)}{|x
   - y|^{D + s_a}} + \int d^D x \frac{\lambda_{abcd}}{4!} \phi (x)^a \phi
   (x)^b \phi (x)^c \phi (x)^d, \]
we can take
\[ s_a = \frac{D + \gamma_a \delta}{2} . \]
We normalize $\gamma_1 = 1$. For local CFTs, from the conformal Wald identity,
we get the following OPE formula (see for example
{\cite{Penedones:2016voo}}),
\[ C_{O O T_{\mu \nu}} = - \frac{D \Delta_O}{D - 1} \frac{1}{S_D}, \]
with $S_D = \frac{2 \pi^{d / 2}}{\Gamma (d / 2)}$. Here $O$ can be any scalar
operator. Apply this to the $\phi_a$'s, we have the OPE ratios
\begin{equation}
  C_{\phi_a \phi_a T_{\mu \nu}} / C_{\phi_b \phi_b T_{\mu \nu}} =
  \frac{\Delta_a}{\Delta_b} . \label{OPEratios}
\end{equation}
These relations, together with $\Delta_{T_{\mu \nu}} = D$, give us $N$
constraints, and allow us to solve for $\delta$ together with $\gamma_a$'s.
One can choose to impose these constraints directly at the $\delta \rightarrow
0$ limit, treating $\gamma_a = 1 + \mathcal{O} (\delta)$ as a perturbative
series. Recently, an efficient method to evaluate the OPE in the long-range
perturbation limit has been developed {\cite{Behan:2023ile}}, which will be
useful in the above project\footnote{We thank Connor Behan for explaining
this.}. A special class of the irreducible fixed points, with $O(N)\times Z_2$  symmetry was recently shown to have spontaneous symmetry breaking at all temperature\cite{Chai:2020onq, Chai:2020zgq}, at least perturbatively in $4-\epsilon$ expansion. A long-range generalization of the models was studied in \cite{Chai:2021djc}. It will be interesting to study these models using our method. 

\

2. {{the Gross-Neveu-Yukawa theory.}} Similar to their short-range
cousins, the long-range Gross-Neveu models also have UV completions given by
the long-range Gross-Neveu-Yukawa models (see {\cite{Chai:2021wac}} for the
test of such a UV completion). The short-range Gross-Neveu-Yukawa fixed point
lives in a long-range conformal manifold parametrized $\Delta_{\psi}$ and
$\Delta_{\phi}$, the conformal manifold contains various perturbative limits,
one therefore has the freedom to choose where to perform the perturbative
calculation. We can consider the couplings
\[ \lambda_1 \overline{\psi_i} \psi_i \phi + \lambda_2 \phi^4, \]
and introduce long-range kinetic terms for both the fermions and bosons. Take
$\Delta_{\psi} = \frac{3 D}{8} + \gamma_{\psi} \delta$ and $\Delta_{\phi} =
\frac{D}{4} + \gamma_{\phi} \delta$, and treat $\delta$ as the perturbation
parameter. Similar to the bosonic theory, we can normalize $\gamma_{\psi} =
1$. The condition $\Delta_T = D$ and the OPE relations similar to
\eqref{OPEratios} allow us to fix $\delta$ and $\gamma_{\phi}$, and recover
the data for the local/short-range Gross-Neveu-Yukawa model. Alternatively,
one can keep only the Yukawa coupling (assume we have tune $\lambda_2 = 0$)
\[ \lambda_1 \overline{\psi_i} \psi_i \phi, \]
and take $\Delta_{\psi} = \frac{1}{6} (2 D + 1) + \gamma_{\psi} \delta$ and
$\Delta_{\phi} = \frac{1}{3}  (D - 1) + \gamma_{\phi} \delta$. In this limit,
the two candidate spin-2 operators, $T^{(f) \mu \nu} = \bar{\psi}
\partial_{\nu} \gamma_{\mu} \psi - \frac{1}{d} \delta_{\mu \nu} \bar{\psi}
\partial \cdot \gamma \psi$ and $T^{(b) \mu \nu} = (\partial^{\mu} \phi
\partial^{\nu} \phi - y \phi \partial^{\mu} \partial^{\nu} \phi -
\tmop{trace})$ are degenerate. They will mix in the perturbation theory.
Compared to the previous setup, the OPE ratios can now be imposed directly at
the $\delta \rightarrow 0$ limit. It is however important to check that the
$\phi^4$ coupling is irrelevant when $\Delta_{T_{\mu \nu}} (\delta) = D$. The
Gross-Neveu-Ising model describes the phase transition from Dirac-semi metal
phase to charge density wave phase on graphene. Other models such as the
Gross-Neveu-Yukawa-Heinsberg and the Gross-Neveu-Yukawa-O(2) model can be
studied in a similar manner. We leave this for future work.

3. {{the quantum O(N) models.}} In addition to the long-range models we
studied here, it is also interesting to consider the quantum long-range
models. For example, the local O(N) vector model can be embedded into a
critical manifold, which can be reached as the IR fixed point of the
following action,
\begin{eqnarray}
 S = &&\int d t d^{D - 1} x  \frac{1}{2} \sum_i (\partial_t \phi^i)^2 + \int
   d t d^{D - 1} x d^{D - 1} y \frac{1}{2} \frac{\sum_i \phi^i (t, x) \phi^i
   (t, y)}{| x - y |^{D - 1 + \sigma}} \nonumber\\&&+ \lambda \int d t d^{D - 1} x \left(
   \sum_i \phi^i (t, x)^2 \right)^2 .
\end{eqnarray}
Some of these models have recently been studied using quantum Monte Carlo
simulation {\cite{Song:2023omz}} and perturbatively in
{\cite{Benedetti:2024oif}}. It will be interesting to use the above theory as
the perturbation theory to study the short-range models by imposing the
conformal Ward identity.

It will also be interesting to generalize our work to the long-range
generalization of the non-Abelian Thirring model
{\cite{frishman1993bosonization,Bennett:1999he,banks1976bosonization,ludwig20034}},
the long-range generalization of QED as studied in
{\cite{di20193d,heydeman2020renormalization}}, and the long-range non-linear
sigma model
{\cite{Brezin:1975sq,hikami1981three,friedan1985nonlinear,sak1977low,bernreuther1986four}}.

\acknowledgments
The author thanks Rajeev Erramilli, Connor Behan, Jiaxin Qiao, Slava Rychkov,
  Edoardo Lauria, \ Philine van Vliet, Bilal Hawashin, Michael Scherer, Tom
  Steudtner and Emmanuel Stamou for stimulating discussions. The author also thanks Zhengwen Liu for introducing him to the Mellin-Barnes method.
 The work of JR is
  supported by the Simons programme g{\'e}n{\'e}ral (2022-2031) in Institut
  des Hautes {\'E}tudes Scientifiques. The author is currently affiliated with Centre de physique th\'eorique, Ecole Polytechnique.

\appendix
\section{spin-2 and spin-1 operator of the scalar theory}
\begin{figure}[h]
\begin{center}
  \resizebox{0.3\columnwidth}{0.2\columnwidth}{\includegraphics{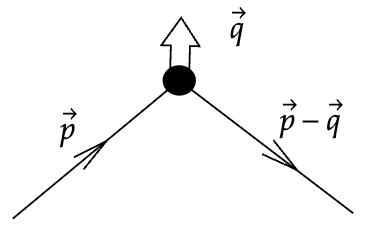}}
\end{center}
  \caption{for calculating $\Delta_T$ and $\Delta_J$.\label{treeleveldiagram}}
\end{figure}

Our calculation is performed using AMBRE and MB.m, which use mostly minus
signature, the change of convention is simply given by
\[ I^{\tmop{Lorentz}, \tmop{most} \tmop{plus}} = \frac{1}{i^L} \frac{1}{(-
   1)^{\nu_1 + \nu_2 + \cdots}} \frac{1}{(2^D \pi^{D / 2})^L} I^{\tmop{AMBRE}}
   . \]
We follow {\cite{Benedetti:2020rrq}} and use the zero momentum BPHZ
(Bogoliubov-Parasiuk-Hepp-Zimmermann) subtraction scheme
{\cite{sibold2010bogoliubov}}. The tree-level diagram is given in Fig.
\ref{treeleveldiagram}, the corresponding Feynman rule for the stress-tensor
vertex is
\[ i \lambda_{\rm ST} ((p - q)_{\mu} ((p - q)_{\nu} - y p_{\nu}) + p_{\mu} (p_{\nu}
   - y (p - q)_{\nu}) - \tmop{trace}) \delta_{i j}, \]
with $y = - \frac{\Delta_{\phi} + 1}{\Delta_{\phi}}$. The Feynman rule for
conserved current vertex is
\[ i \lambda_J (p_{\mu} + (p - q)_{\mu}) \Omega_{i j} . \]

\begin{figure}[h]
\begin{center}
  \resizebox{1.5in}{2in}{\includegraphics{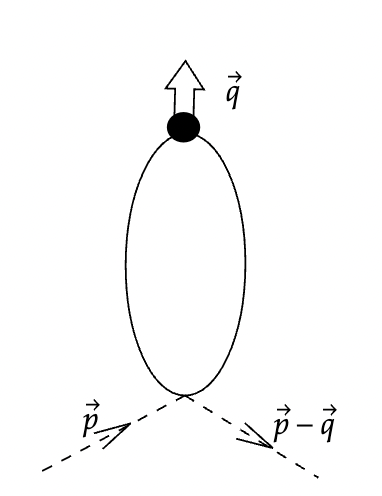}}
\end{center}
  \caption{One loop diagram for calculating $\Delta_T$ and
  $\Delta_J$.\label{oneloopdiagram}}
\end{figure}

In the leading order, the diagram which contributes to the renormalization of
the stress tensor is given in Fig. \ref{oneloopdiagram}.

After introducing the IR regulator for our propagator, we can perform the
regularization at any kinematic point. For simplicity, we take
\[ q^{\mu} = 0. \]
In this special kinematic point, the diagram must vanish. Since there is no
external momentum that one can use to build an expression that transforms in
the spin-2 representation of the Euclidean group. We therefore conclude that
this diagram must be regular. This trick was introduced in
{\cite{Behan:2019lyd}}.

\begin{figure}[h]
\begin{center}
  \resizebox{4in}{!}{\includegraphics{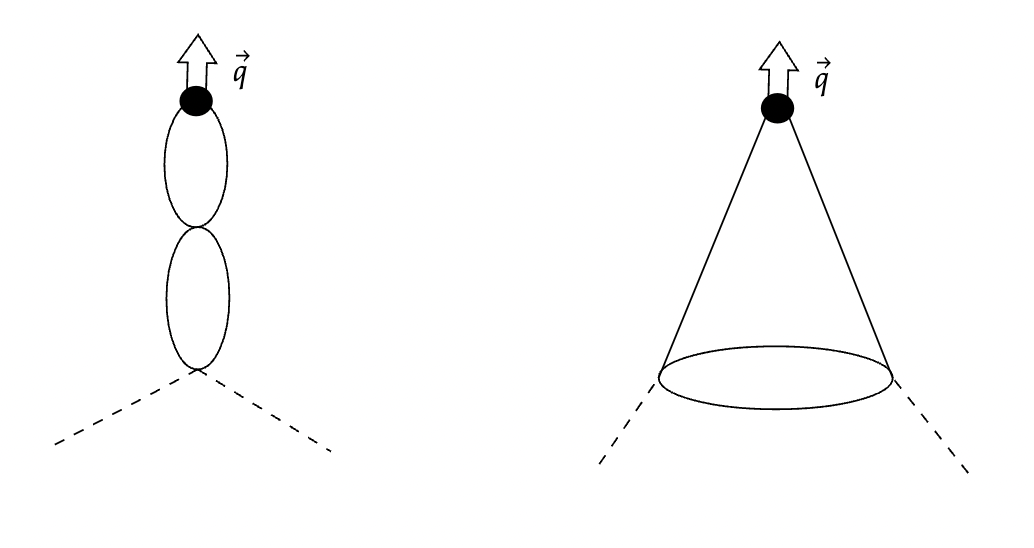}}
  \caption{Two loop diagram for calculating $\Delta_T$ and
  $\Delta_J$..\label{twoloop}}
\end{center}
\end{figure}

The two-loop diagrams are given in Fig. \ref{twoloop}.

We again renormalize our diagram at the special kinematic point
\[ q^{\mu} = 0. \]
By the same argument as before, the first diagram should vanish at $q^{\mu} =
0$ limit. The second diagram leads us to the integral
\begin{eqnarray}
 \mu^{2 \delta} \mathcal{I}_0^{\mu \nu} = && \frac{1}{((2 \pi)^D)^2} \int d
   k_1^D d k_2^D \frac{k_2^{\mu} k_2^{\nu} - \tmop{trace}}{(k_2 ^2 + \mu^2)^{s
   / 2} ((p - k_1  + k_2)^2 + \mu^2)^{s / 2} (k_1 ^2 + \mu^2)^{s / 2} (k_1 ^2
   + \mu^2)^{s / 2}} \nonumber\\=&& (p^{\mu} p^{\nu} - \tmop{trace}) S_2, 
\end{eqnarray}
 Using AMBRE and then MB.m, we get
\begin{eqnarray}
  S_2 =&& \frac{(4 \pi)^{- D}}{\delta \Gamma \left( \frac{D}{2} + 2 \right)
  \Gamma \left( \frac{D}{2} \right)} + \frac{(4 \pi)^{- D}}{\Gamma \left(
  \frac{D}{2} + 2 \right) \Gamma \left( \frac{D}{4} \right)^2 \Gamma \left(
  \frac{D}{2} \right)} J_2 (D) \nonumber\\&& - \frac{2^{- 2 D - 1} \pi^{- D} }{\Gamma \left(
  \frac{D}{2} + 2 \right) \Gamma \left( \frac{D}{2} \right)} \left( \psi^{(0)}
  \left( \frac{D}{2} \right) + 2 \psi^{(0)} \left( \frac{D}{4} \right) + 3
  \gamma - 1 \right) . \label{s2constant}
\end{eqnarray}
Here we introduce a constant $J_2 (D)$, whose analytical expression are not
easy to calculate, we instead give the corresponding Mellin-Barnes integral,
\[ J_2 (D) = \int_{0^{^-}} d z \frac{\Gamma (2 - z) \Gamma (- z) \Gamma (z)
   \Gamma \left( \frac{D}{4} + z \right)^2 \Gamma \left( \frac{D}{2} + z
   \right)}{\Gamma \left( \frac{D}{2} + 2 z \right)} . \]
The integral is regular at $D = 4$. The best way to evaluate is to numerically
integrate along the contour (after shifting the contour away from poles),
which can be easily done in MB.m. We also give an approximate result of the
integral in $4 - \epsilon$,
\[ J_2 (4 - \epsilon) = - 2.229271492460209 - 1.077724442678504 \epsilon -
   0.5458130816580664 \epsilon^2 + O (\epsilon^3) . \]
Also, notice
\[ J_2 (D = 3) \approx - 4.1944. \]
In our calculation of the O(N) model, this constant will drop out in the final
expression, it is, however, important to consider it for other multi-component
scalar models.

\begin{figure}[h]
\begin{center}
  \resizebox{4in}{!}{\includegraphics{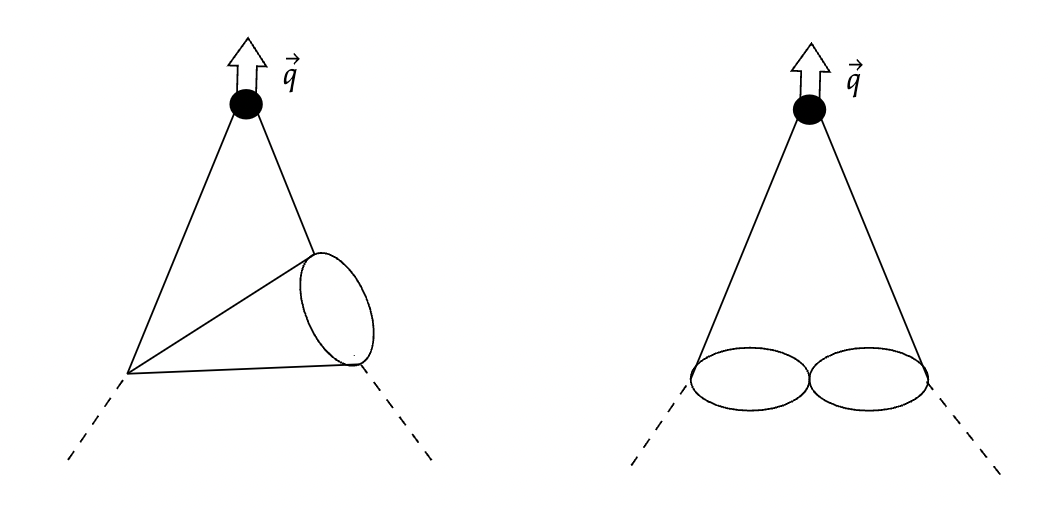}}
\end{center}
  \caption{The three loop diagram for $\Delta_T$ and
  $\Delta_J$.\label{threeloop}}
\end{figure}

At three-loop level, the diagrams that survive after taking \ $q^{\mu} = 0$
are given in Fig. \ref{threeloop}.
The corresponding integrals are
\begin{eqnarray*}
  \mu^{2 \delta} \mathcal{I}_1^{\mu \nu} & = & \frac{\mu^{3 \delta}}{((2
  \pi)^D)^3} \int d^D k_1 d^D k_2 d^D k_3\\
  &  & \times \frac{k_3^{\mu} k_3^{\nu} - \tmop{trace}}{(k_3 ^2 + \mu^2)^s
  ((p + k_2)^2 + \mu^2)^{s / 2} ((k_1 + k_2) ^2 + \mu^2)^{s / 2} ((k_2 + k_3)
  ^2 + \mu^2)^{s / 2} (k_1 ^2 + \mu^2)^{s / 2}}\\
  & = & (p^{\mu} p^{\nu} - \tmop{trace}) I_1
\end{eqnarray*}
\begin{eqnarray*}
  \mu^{2 \delta} \mathcal{I}_2^{\mu \nu} & = & \frac{\mu^{3 \delta}}{((2
  \pi)^D)^3} \int d^D k_1 d^D k_2 d^D k_3\\
  &  & \times \frac{k_3^{\mu} k_3^{\nu} - \tmop{trace}}{(k_3 ^2 + \mu^2)^{s /
  2} ((p + k_1 - k_3)^2 + \mu^2)^{s / 2} ((k_2) ^2 + \mu^2)^{s / 2} ((p + k_2
  - k_3) ^2 + \mu^2)^{s / 2} (k_3 ^2 + \mu^2)^s}\\
  & = & (p^{\mu} p^{\nu} - \tmop{trace}) I_2
\end{eqnarray*}
with
\begin{eqnarray}
  I_1 & = & \frac{2^{2 - 3 D} \pi^{- \frac{3 D}{2}}}{3 \delta^2 \Gamma \left(
  \frac{D}{2} + 2 \right) \Gamma \left( \frac{D}{2} \right)^2} \nonumber\\
  &  & + \frac{2^{1 - 3 D} \pi^{- \frac{3 D}{2}}  \left( 3 J_0 (D) - \Gamma
  \left( \frac{D}{4} \right)^2  \left( 5 \psi^{(0)} \left( \frac{D}{4} \right)
  + \psi^{(0)}  \left( \frac{D}{4} + 2 \right) + 6 \gamma - 2 \right)
  \right)}{3 \delta \Gamma \left( \frac{D}{2} + 2 \right) \Gamma \left(
  \frac{D}{4} \right)^2 \Gamma \left( \frac{D}{2} \right)^2} \nonumber\\
  I_2 & = & \frac{8^{1 - D} \pi^{- \frac{3 D}{2}}}{3 \delta^2 \Gamma \left(
  \frac{D}{2} + 2 \right) \Gamma \left( \frac{D}{2} \right)^2} + \frac{2^{2 -
  3 D} \pi^{- \frac{3 D}{2}} }{3 \delta \Gamma \left( \frac{D}{2} + 2 \right)
  \Gamma \left( \frac{D}{4} \right)^2 \Gamma \left( \frac{D}{2} \right)^2}
  \bigg( 3 J_0 (D) \nonumber\\&& - \Gamma \left( \frac{D}{4} \right)^2  \left( \psi^{(0)}
  \left( \frac{D}{2} \right) + 4 \psi^{(0)} \left( \frac{D}{4} \right) + 5
  \gamma - 2 \right) \bigg) . \label{i2constant} 
\end{eqnarray}
Collect the results, we get the renormalization function of the stress-tensor
vertex (at $q^{\mu} = 0$) $$\Gamma^{\mu \nu} = 2 (p^{\mu} p^{\nu} -
\tmop{trace}) \Gamma_{\rm ST},$$ with
\[ \Gamma_{\rm ST} = \lambda_{\rm ST} \left( 1 + \frac{3}{2} \lambda^2 (N + 2) S_2 \mu^{- 2
   \delta} - I_1 \lambda^3  (N^2 + 10 N + 16) \mu^{- 3 \delta} - \frac{1}{4}
   I_2 \lambda^3  (N^2 + 10 N + 16) \mu^{- 3 \delta} \right) . \]
We define our renormalized coupling to be
\begin{equation}
  g_{\rm ST} = \mu^{- (D - 2 \Delta_{\phi} - 2)} \Gamma_{\rm ST} .
\end{equation}
Invert the above relation, we get
\begin{eqnarray}
  \lambda_{\rm ST} = &&\mu^{(D - 2 \Delta_{\phi} - 2)} g_{\rm ST} \bigg( 1 - \frac{3}{2}
  \lambda^2 (N + 2) S_2 \mu^{- 2 \delta} \nonumber\\&&+ \frac{1}{4} \lambda^3 \mu^{- 3
  \delta}  (4 I_2 (N^2 + 10 N + 16) + I_2 (N^2 + 10 N + 16)) + 1 \bigg)
  \label{renormalizeLT}
\end{eqnarray}
The beta function for the coupling $g_{\rm ST}$ is defined as
\[ \beta_{\rm ST}= \mu \frac{\partial}{d \mu} g_{\rm ST} . \]
Notice the bare couplings $\lambda_{\rm ST}$ and $\lambda$ do not depend on $\mu$,
so that the above definition together with $\eqref{renormalizeLT}$ and
\begin{equation}
  \lambda = \mu^{\delta}  \left( g + \frac{1}{2} D_0 g^2 (N + 8) + \frac{1}{4}
  D_0^2 g^3  (N^2 + 6 N + 20) + 3 g^3  \left( \frac{5 N}{3} + \frac{22}{3}
  \right)  \left( \text{$D_0$}^2 - S_0 \right) \right), \label{invertG}
\end{equation}
gives us the scaling dimension of the (to-be) stress tensor, which is
\eqref{spin2ing}. The constants $D_0$ and $S_0$ are defined in
\eqref{DScoupling}. The above relation comes from the renormalization of the
coupling constant, which was calculated in {\cite{Benedetti:2020rrq}}.

One can similarly calculate the scaling dimension of the current operator
\eqref{spin1ing}. The diagrams that we need to calculate are the same as the
ones for the stress tensor. We can also use the special kinematics $q^{\mu} =
0$, which means that the one-loop diagram in Fig. \ref{oneloopdiagram} and the
first two-loop diagram in Fig. \ref{twoloop} do not contribute to the final
result. The two-loop integral corresponding to the second diagram in Fig.
\ref{twoloop} \ is
\begin{eqnarray}
 \mu^{2 \delta} {\mathcal{I}'_0}^{\mu \nu} = &&\frac{1}{((2 \pi)^D)^2} \int
   d^D k_1 d^D k_2 \frac{k_1^{\mu}}{(k_2 ^2 + \mu^2)^{s / 2} ((p - k_1  +
   k_2)^2 + \mu^2)^{s / 2} (k_1 ^2 + \mu^2)^{s / 2} (k_1 ^2 + \mu^2)^{s / 2}}
   \nonumber\\&&= p^{\mu} S_1,
\end{eqnarray}
with
\begin{eqnarray}
  S_1 =&& \frac{(4 \pi)^{- D}}{\delta \Gamma \left( \frac{D}{2} + 1 \right)
  \Gamma \left( \frac{D}{2} \right)} + \frac{(4 \pi)^{- D}}{\Gamma \left(
  \frac{D}{2} + 1 \right) \Gamma \left( \frac{D}{4} \right)^2 \Gamma \left(
  \frac{D}{2} \right)} J_1 (D) \nonumber\\&& - \frac{2^{- 2 D - 1} \pi^{- D} }{\Gamma \left(
  \frac{D}{2} + 1 \right) \Gamma \left( \frac{D}{2} \right)} \left( \psi^{(0)}
  \left( \frac{D}{2} \right) + 2 \psi^{(0)} \left( \frac{D}{4} \right) + 3
  \gamma \right), \label{s1constant}
\end{eqnarray}
and
\[ J_1 (D) = \int_{0^{^-}} d z \frac{\Gamma (1 - z) \Gamma (- z) \Gamma (z)
   \Gamma \left( \frac{D}{4} + z \right)^2 \Gamma \left( \frac{D}{2} + z
   \right)}{\Gamma \left( \frac{D}{2} + 2 z \right)} . \]
After numerical evaluation, we have $J_1 (3) \approx - 3.08557$. The integral
is regular at $D = 4$. We can give an expression in $4 - \epsilon$,
\begin{eqnarray}
J_1 (4 - \epsilon) \approx && - 1.075405220902442 \epsilon^4 -
   1.401988434885228 \epsilon^3 - 1.697234636355991 \epsilon^2\nonumber\\&& -
   1.640421373262027 \epsilon - 1.562604825793316.
\end{eqnarray}
In our calculation of the O(N) model, this constant again drops out in the
final expression, it is however, important to consider it for other
multi-component scalar models.
The three-loop integrals corresponding to the diagrams in Fig.
\ref{threeloop} are
\begin{eqnarray*}
  \mu^{2 \delta} {\mathcal{I}'_1}^{\mu \nu} & = & \frac{\mu^{3 \delta}}{((2
  \pi)^D)^3} \int d^D k_1 d^D k_2 d^D k_3\\
  &  & \times \frac{k_3^{\mu}}{(k_3 ^2 + \mu^2)^s ((p + k_2)^2 + \mu^2)^{s /
  2} ((k_1 + k_2) ^2 + \mu^2)^{s / 2} ((k_2 + k_3) ^2 + \mu^2)^{s / 2} (k_1 ^2
  + \mu^2)^{s / 2}}\\
  & = & p^{\mu} I_1'
\end{eqnarray*}
\begin{eqnarray*}
  \mu^{2 \delta} {\mathcal{I}'_2}^{\mu \nu} & = & \frac{\mu^{3 \delta}}{((2
  \pi)^D)^3} \int d^D k_1 d^D k_2 d^D k_3\\
  &  & \times \frac{k_3^{\mu}}{(k_3 ^2 + \mu^2)^{s / 2} ((p + k_1 - k_3)^2 +
  \mu^2)^{s / 2} ((k_2) ^2 + \mu^2)^{s / 2} ((p + k_2 - k_3) ^2 + \mu^2)^{s /
  2} (k_3 ^2 + \mu^2)^s}\\
  & = & p^{\mu} I_4
\end{eqnarray*}
The constants are
\begin{eqnarray}
  I_1' =&& \frac{2^{2 - 3 D} \pi^{- \frac{3 D}{2}}}{3 \delta^2 \Gamma \left(
  \frac{D}{2} + 1 \right) \Gamma \left( \frac{D}{2} \right)^2} + \frac{2^{1 -
  3 D} \pi^{- \frac{3 D}{2}} }{3 \delta \Gamma \left( \frac{D}{2} + 1 \right)
  \Gamma \left( \frac{D}{4} \right)^2 \Gamma \left( \frac{D}{2} \right)^2}
  \bigg[  3 J_1 (D)  \nonumber\\&&- \Gamma \left( \frac{D}{4} \right)^2  \left( 5 \psi^{(0)} \left(
  \frac{D}{4} \right) + \psi^{(0)}  \left( \frac{D}{4} + 1 \right) + 6 \gamma
  \right) \bigg] . \label{i1pconstant}
\end{eqnarray}
and
\begin{eqnarray}
  I_2' = && \frac{8^{1 - D} \pi^{- \frac{3 D}{2}}}{3 \delta^2 \Gamma \left(
  \frac{D}{2} + 1 \right) \Gamma \left( \frac{D}{2} \right)^2} + \frac{2^{2 -
  3 D} \pi^{- \frac{3 D}{2}} }{3 \delta \Gamma \left( \frac{D}{2} + 1 \right)
  \Gamma \left( \frac{D}{4} \right)^2 \Gamma \left( \frac{D}{2} \right)^2}
  \bigg[ 3 J_1 (D) \nonumber\\&&- \Gamma \left( \frac{D}{4} \right)^2  \left( \psi^{(0)}
  \left( \frac{D}{2} \right) + 4 \psi^{(0)} \left( \frac{D}{4} \right) + 5
  \gamma \right) \bigg] . \label{i2pconstant}
\end{eqnarray}
With the above integral, we can calculate the renormalization of the conserved
current vertex, which gives $\Gamma^{\mu} = 2 p^{\mu} \Gamma_J$, with
\[ \Gamma_J = \lambda_J \left( 1 + + \frac{1}{2} \lambda^2 (N + 2) S_2 \mu^{-
   2 \delta} - 2 I_1' \lambda^3 (N + 2) \mu^{- 3 \delta} - \frac{1}{4} I_2'
   \lambda^3 (N^2 + 6 N + 8) \mu^{- 3 \delta} \right) . \]
We define our renormalized coupling to be
\[ g_J = \mu^{- (D - 2 \Delta_{\phi} - 1)} \Gamma_J . \]
Invert the above relation, we get
\begin{equation}
  \lambda_J = \mu^{(D - 2 \Delta_{\phi} - 1)} g_J \left( 1 - - \frac{1}{2}
  \lambda^2 (N + 2) S_2' \mu^{- 2 \delta} + \frac{1}{4} \lambda^3 \mu^{- 3
  \delta}  (8 I_1' (N + 2) + I_2' (N^2 + 6 N + 8)) \right)
  \label{renormalizeLJ}
\end{equation}
Together with \eqref{invertG} and the definition $\beta_J = \mu
\frac{\partial}{d \mu} g_J$, the above results give us the scaling dimension
of the (to-be) conserved current, which is \eqref{spin1ing}.

\

\section{Fermionic models in $D = 2 + 1$ and $D = 4 +
1$}\label{thirringdetails}
\begin{figure}[h]
\begin{center}
  \resizebox{2in}{1in}{\includegraphics{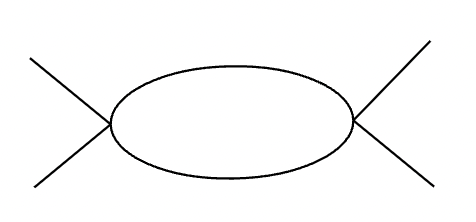}}
\end{center}  
  \caption{The one-loop diagram for the beta function of the fermionic models.
  \label{betafermionic}}
\end{figure}

\begin{figure}[h]
\begin{center}
  \resizebox{3in}{!}{\includegraphics{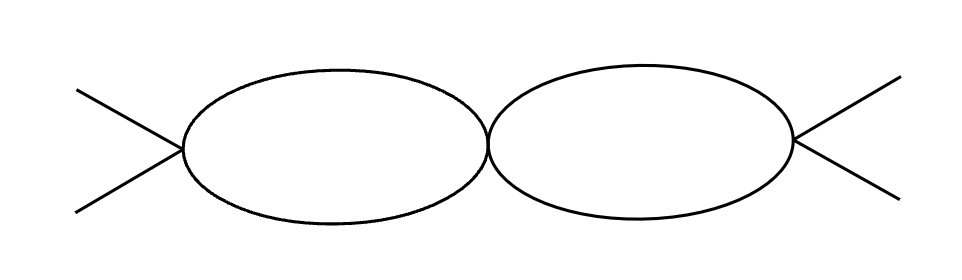}}\resizebox{2in}{!}{\includegraphics{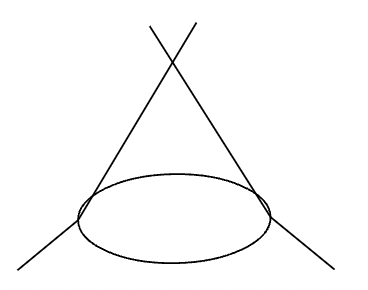}}
\end{center} 
  \caption{The two-loop diagrams for the beta function of the fermionic
  models. \label{betafermionic2loop}}
\end{figure}

The one and two-loop diagrams which contribute to the beta function
renormalization are given in Fig. \ref{betafermionic} and Fig
\ref{betafermionic2loop}.

The diagrams contributing to the renormalization of $T_{\mu\nu}$ have the same topology as the bosonic case. The Feynman rule for the stress-tensor vertex is
$$ i \lambda_{\rm ST}( (\epsilon . \gamma^{\mu})_{\alpha \beta} (p^{\nu}
   + (p - q)^{\nu}) + (\epsilon . \gamma^{\nu})_{\alpha \beta} (p^{\mu} + (p -
   q)^{\mu} - {\rm trace}) \delta_{a b} $$
for the $D = 2 + 1$ model \eqref{3dfermionaction}. For the $D = 4 + 1$
dimensional model \eqref{5dThirring}, on the other hand, the Feynman rule is
$$ i \lambda_{\rm ST}( (C. \gamma^{\mu})_{\alpha \beta} (p^{\nu} + (p -
   q)^{\nu}) + (C. \gamma^{\nu})_{\alpha \beta} (p^{\mu} + (p - q)^{\mu} -
   {\rm trace}) \Omega_{i j} .$$
We will first summarize the integrals that we will encounter in our
calculation. After evaluating the $\gamma$-matrix algebra, we are left with
the following integrals, which we again evaluate using the Mellin-Barnes
method.

At one loop, we will encounter two integral. The first integral is
$$
\mu^\delta \times I_0^{\mu \nu}=\frac{1}{(2 \pi)^D} \int d^D k \frac{k^\mu k^\nu}{\left(k^2+\mu^2\right)^{s / 2}\left(k^2+\mu^2\right)^{s / 2}}=f_1 \eta^{\mu \nu}=-\frac{2^{-D-1} \pi^{-\frac{D}{2}} \cdot \Gamma\left(\frac{\delta}{2}\right)}{\Gamma\left(\frac{1}{2} \cdot(D+\delta+2)\right)} \eta^{\mu \nu} .
$$

The second integral is
$$
\mu^\delta \times I_1^{\mu \nu}=\frac{1}{(2 \pi)^D} \int d^D k \frac{1}{\left(k^2+\mu^2\right)^{s / 2}\left(k^2+\mu^2\right)^{s / 2}}=\frac{1}{{\mu}^2} f_2=\frac{1}{{\mu}^2} \frac{2^{-D} \cdot \pi^{-\frac{D}{2}} \cdot \Gamma\left(\frac{\delta}{2}+1\right)}{\Gamma \left(\frac{1}{2} \cdot(D+\delta+2)\right)} .
$$

At two loop, we need to evaluate
$$
\begin{aligned}
& \mu^{2 \delta} I_3^{\mu \nu}=\frac{1}{\left((2 \pi)^D\right)^2} \int d^D k_1 d^D k_2 \frac{k_2^\mu k_2^\nu}{\left(k_2^2+\mu^2\right)^{s / 2}\left(\left(p-k_1+k_2\right)^2+\mu^2\right)^{s / 2}\left(k_1^2+\mu^2\right)^{s/2}\left(k_1^2+\mu^2\right)^{s / 2}}= \\
& \frac{1}{\delta} \eta^{\mu \nu} c_3+\mathcal{O}\left(\delta^0\right)
\end{aligned}
$$
Here the constant is $c_3=-\frac{(4 \cdot \pi)^{-D}}{\mu^2 \cdot \Gamma \cdot\left(\frac{D}{2}+1\right)^2}$. The next integral is
$$
\begin{aligned}
\mu^{2 \delta} I_5^{\mu_1 \mu_2 \mu_3 \mu_4} & =\frac{1}{\left((2 \pi)^D\right)^2} \int d^D k_1 d^D k_2 \frac{k_1^{\mu_1} k_1^{\mu_2} k_1^{\mu_3} k_2^{\mu_4}}{\left(k_2^2+\mu^2\right)^{s / 2}\left(\left(p-k_1+k_2\right)^2+\mu^2\right)^{s / 2}\left(k_1^2+\mu^2\right)^{s / 2}\left(k_1^2+\mu^2\right)^{s / 2}} \\
& =\frac{1}{\delta}\left(\eta^{\mu_1 \mu_2} \eta^{\mu_3 \mu_4}+\eta^{\mu_1 \mu_3} \eta^{\mu_2 \mu_4}+\eta^{\mu_1 \mu_4} \eta^{\mu_2 \mu_3}\right) c_5+\mathcal{O}\left(\delta^0\right)
\end{aligned}
$$
with the constant is $c_5=\frac{4^{-D-1} \cdot \pi^{-D} \cdot \Gamma\left(\frac{D-2}{4}\right)}{\Gamma \cdot\left(\frac{D}{2}+2\right) \cdot \Gamma\left(\frac{D}{2}\right) \cdot \Gamma\left(\frac{D+2}{4}\right)}$. The next integral is
$$
\begin{aligned}
\mu^{2 \delta} I_6^{\mu_1 \mu_2 \mu_3 \mu_4} & =\frac{1}{\left((2 \pi)^D\right)^2} \int d^D k_1 d^D k_2 \frac{k_1^{\mu_1} k_1^{\mu_2} k_2^{\mu_3} k_2^{\mu_4}}{\left(k_2^2+\mu^2\right)^{s / 2}\left(\left(p-k_1+k_2\right)^2+\mu^2\right)^{s / 2}\left(k_1^2+\mu^2\right)^{s / 2}\left(k_1^2+\mu^2\right)^{s / 2}} \\
& =\frac{1}{\delta^2} \eta^{\mu_1 \mu_2} \eta^{\mu_3 \mu_4} c_{6,0}+\frac{1}{\delta} \eta^{\mu_1 \mu_2} \eta^{\mu_3 \mu_4} c_{6,1}+\frac{1}{\delta}\left(\eta^{\mu_1 \mu_3} \eta^{\mu_2 \mu_4}+\eta^{\mu_1 \mu_4} \eta^{\mu_2 \mu_3}\right) c_{6,2}+\mathcal{O}\left(\delta^0\right)
\end{aligned}
$$
with the constants
$$
\begin{aligned}
& c_{6,0}=\frac{2^{-2 D-1} \cdot \pi^{-D}}{\Gamma \left(\frac{D}{2}+1\right)^2} \\
& c_{6,1}=\frac{(4 \cdot \pi)^{-D} \cdot\left((2-D) \cdot {\rm Ha}\left(\frac{D}{2}\right)-2 \cdot(D-2) \cdot {\rm Ha}\left(\frac{D-2}{4}\right)+2\right)}{(D-2) \cdot D^2 \cdot \Gamma\left(\frac{D}{2}\right)^2} \\
& c_{6,2}=\frac{2^{-2 \cdot D-1} \cdot \pi^{-D}}{(D-2) \cdot \Gamma \left(\frac{D}{2}+1\right)^2} .
\end{aligned}
$$
Here $\mathrm{Ha}(x)$ is the Harmonic number. Next, we have
$$
\begin{aligned}
\mu^{2 \delta} I_7^{\mu_1 \mu_2 \mu_3 \mu_4} & =\frac{1}{\left((2 \pi)^D\right)^2} \int d^D k_1 d^D k_2 \frac{k_1^{\mu_1} k_1^{\mu_2} k_1^{\mu_3} k_1^{\mu_4}}{\left(k_2{ }^2+\mu^2\right)^{s / 2}\left(\left(p-k_1+k_2\right)^2+\mu^2\right)^{s / 2}\left(k_1{ }^2+\mu^2\right)^{s / 2}\left(k_1{ }^2+\mu^2\right)^{s / 2}} \\
& =\frac{1}{\delta}\left(\eta^{\mu_1 \mu_2} \eta^{\mu_3 \mu_4}+\eta^{\mu_1 \mu_3} \eta^{\mu_2 \mu_4}+\eta^{\mu_1 \mu_4} \eta^{\mu_2 \mu_3}\right) c_7+\mathcal{O}\left(\delta^0\right)
\end{aligned}
$$
with the constants $c_7=\frac{2^{-2 D-1} \cdot \pi^{-D}}{(D-2) \cdot \Gamma \left(\frac{D}{2}+1\right)^2}$.
We also encounter three regular integrals that appears in the calculation,
$$
\begin{aligned}
\mu^{2 \delta} I_8 & =\frac{1}{\left((2 \pi)^D\right)^2} \int d^D k_1 d^D k_2 \frac{1}{\left(k_2^2+\mu^2\right)^{s / 2}\left(\left(p-k_1+k_2\right)^2+\mu^2\right)^{s / 2}\left(k_1^2+\mu^2\right)^{s / 2}\left(k_1^2+\mu^2\right)^{s / 2}}, \\
\mu^{2 \delta} I_2^{\mu \nu} & =\frac{1}{\left((2 \pi)^D\right)^2} \int d^D k_1 d^D k_2 \frac{k_1^\mu k_1^\nu}{\left(k_2^2+\mu^2\right)^{s / 2}\left(\left(p-k_1+k_2\right)^2+\mu^2\right)^{s / 2}\left(k_1^2+\mu^2\right)^{s / 2}\left(k_1^2+\mu^2\right)^{s / 2}}, \\
\mu^{2 \delta} I_4^{\mu \nu} & =\frac{1}{\left((2 \pi)^D\right)^2} \int d^D k_1 d^D k_2 \frac{k_1^\mu k_2^\nu}{\left(k_2^2+\mu^2\right)^{s / 2}\left(\left(p-k_1+k_2\right)^2+\mu^2\right)^{s / 2}\left(k_1^2+\mu^2\right)^{s / 2}\left(k_1^2+\mu^2\right)^{s / 2}}.
\end{aligned}
$$
For the $2+1 \mathrm{D}$ fermionic model, we can calculate the renormalization functions (the one-particle irreducible two/three/four-point functions),
$$
\Gamma_{\lambda_Y}, \quad \Gamma_{\lambda_T}, \quad \Gamma_{\mathrm{ST}}^{\mu \nu} \text {, and } \quad \Gamma_{\text {Mass. }}
$$
The renormalization function for four-fermion interaction is
\begin{eqnarray*}
  \Gamma_{\alpha \beta \gamma \delta, a b c d} & = & \Gamma_{\lambda_Y}
  (\delta_{ad} \epsilon_{\alpha \delta} \delta_{bc} \epsilon_{\beta \gamma} +
  \delta_{ac} \epsilon_{\alpha \gamma} \delta_{bd} \epsilon_{\beta \delta} +
  \delta_{ab} \epsilon_{\alpha \beta} \delta_{cd} \epsilon_{\gamma \delta})\\
  &  & + \Gamma_{\lambda_T} (\Omega_{ad} \Omega_{bc} (\epsilon .
  \gamma^{\mu})_{\beta \gamma}  (\epsilon . \gamma_{\mu})_{\alpha \delta} +
  \Omega_{ac} \Omega_{bd}  (\epsilon . \gamma^{\mu})_{\alpha \gamma} 
  (\epsilon . \gamma_{\mu})_{\beta \delta} + \Omega_{ab} \Omega_{cd} 
  (\epsilon . \gamma^{\mu})_{\gamma \delta}  (\epsilon . \gamma_{\mu})_{\alpha
  \beta})
\end{eqnarray*}
with
\begin{eqnarray*}
  \Gamma_{\lambda_Y} & = & \mu^{- 2 \delta} \lambda_Y^3  \big( (6 - 18 N)
  c_{6, 0} - 18 Nc_{6, 1} - 12 Nc_{6, 2} + 6 c_{6, 1} - 36 c_{6, 2} + 18 c_3
  \mu^2 N\\&&  - 30 c_3 \mu^2 - 30 c_5 (N + 1) + 9 f_1^2 N^2 + f_2^2 N^2 - 6 f_1
  f_2 N^2 
  \nonumber\\&& - 27 f_1^2 N - 3 f_2^2 N + 18 f_1 f_2 N + 33 f_1^2 + 5 f_2^2 - 18
  f_1 f_2 \big) 
  \\&&- 6 \mu^{- 2 \delta} \lambda_T^2  (2 \lambda_T  (2 (N + 11)
  c_{6, 0} + 2 Nc_{6, 1} + 8 Nc_{6, 2} + 22 c_{6, 1} \nonumber\\&&+ 28 c_{6, 2} + c_3 \mu^2
  (- N) - 8 c_3 \mu^2 + 10 c_5 (N + 5)) \nonumber\\&&+ 2 f_1 \mu^{\delta} - f_2
  \mu^{\delta}) + \mu^{- 2 \delta} \lambda_Y^2  (- 36 \lambda_T  (3 c_{6, 0} +
  3 c_{6, 1} + 2 c_{6, 2} - 3 c_3 \mu^2 + 5 c_5) \nonumber\\&&+ f_2 (N - 4) (-
  \mu^{\delta}) + 3 f_1 (N - 2) (\mu^{\delta} - 18 f_2 \lambda_T) + 81 f_1^2
  (N - 2) \lambda_T + 9 f_2^2 (N - 2) \lambda_T) 
\nonumber  \\&&+ \mu^{- 2 \delta} \lambda_Y 
  \bigg( - 9 \lambda_T^2  \big( 6 Nc_{6, 0} + 6 Nc_{6, 1} + 4 Nc_{6, 2} + 10
  c_{6, 0} + 10 c_{6, 1} + 20 c_{6, 2} + 10 c_5 (N + 3) \nonumber\\&&- 41 f_1^2 + 26 f_2
  f_1 - 5 f_2^2 \big)
 + 18 c_3 \mu^2 (N - 3) \lambda_T^2 + 6 (3 f_1 - f_2)
  \mu^{\delta} \lambda_T + \mu^{2 \delta} \bigg),\\
  \Gamma_{\lambda_T} & = & - 3 \mu^{- 2 \delta} \lambda_T \lambda_Y^2  (6 (N +
  1) c_{6, 0} + 6 Nc_{6, 1} + 4 N c_{6, 2} + 6 c_{6, 1} + 44 c_{6, 2} - 6 c_3
  \mu^2 N + 18 c_3 \mu^2 \nonumber\\&&+ 10 c_5 (N + 5) - 15 f_1^2 - 3 f_2^2 + 6 f_1 f_2) +
  \mu^{- 2 \delta} \lambda_T  \big( \lambda_T^2  \big( 2 Nc_{6, 1} - 52
  Nc_{6, 2} \nonumber\\&&+ 2 (N - 85) c_{6, 0} - 170 c_{6, 1} - 260 c_{6, 2} - 10 c_5 (5 N
  + 43) + f_1^2 N^2 + f_2^2 N^2 \nonumber\\&&+ 2 f_1 f_2 N^2 + 3 f_1^2 N + 3 f_2^2 N + 6
  f_1 f_2 N + 125 f_1^2 + 17 f_2^2 - 74 f_1 f_2 \big) \nonumber\\&&+ 2 c_3 \mu^2 (N + 5)
  \lambda_T^2 + (f_1 + f_2) (N + 2) \mu^{\delta} \lambda_T + \mu^{2 \delta}
  \big) \nonumber\\&&+ 3 \mu^{- 2 \delta} \lambda_T \lambda_Y  \big( - 4 \lambda_T 
  \big( 2 Nc_{6, 0} + 2 Nc_{6, 1} + 8 Nc_{6, 2} + 13 c_{6, 0} + 13 c_{6, 1}\nonumber\\&& +
  22 c_{6, 2} + c_3 \mu^2 N - 7 c_3 \mu^2 + 5 c_5 (2 N + 7) \big) \nonumber\\&&+ 2 f_2
  \mu^{\delta} + f_1  (2 f_2 (N + 2) \lambda_T - 2 \mu^{\delta}) + f_1^2 (N +
  2) \lambda_T + f_2^2 (N + 2) \lambda_T \big),
\end{eqnarray*}
The renormalization function with one stress tensor and two fermions is
\[  (\Gamma^{\mu \nu})_{\alpha \beta, a b} = 2 i \Gamma_{\tmop{ST}} \left(
   p^{\nu}  (\epsilon . \gamma^{\mu})_{\alpha \beta} + p^{\nu}  (\epsilon .
   \gamma^{\mu})_{\alpha \beta} - \frac{2}{3} \eta^{\mu \nu} p^{\sigma} 
   (\epsilon . \gamma^{\sigma})_{\alpha \beta} \right) \delta_{a b}, \]
with
\[ \begin{array}{lll}
     \Gamma_{\tmop{ST}} & = & \lambda_{\tmop{ST}} + 6 c_5  (N + 1) \mu^{- 2
     \delta} \lambda_{\tmop{ST}} \lambda_T^2 + 2 c_5  (N - 1) \mu^{- 2 \delta}
     \lambda_{\tmop{ST}} \lambda_Y^2 + 12 c_5 \mu^{- 2 \delta}
     \lambda_{\tmop{ST}} \lambda_T \lambda_Y .
   \end{array} \]
The renormalization function with one mass operator and two fermions is,
\[ \Gamma_{\alpha \beta, a b} = \Gamma_{\tmop{Mass}} \epsilon_{\alpha \beta}
   \delta_{a b}, \]
with
\begin{eqnarray*}
  \Gamma_{\tmop{Mass}} & = & \lambda_{\tmop{Mass}}  \bigg( 9 \mu^{- 2 \delta}
  \lambda_T^2  \big( - 3 (N + 1) c_{6, 1} - 2 Nc_{6, 2} - 2 c_{6, 2} + c_3
  \mu^2 N + c_3 \mu^2 \nonumber\\&&+ 5 c_5 (N + 1) + 9 f_1^2 + f_2^2 + 6 f_1 f_2 \big) +
  6 \mu^{- 2 \delta} \lambda_T \lambda_Y  \big( 3 (- 3 c_{6, 1} - 2 c_{6, 2}
  + c_3 \mu^2 + 5 c_5) \nonumber\\&&+ \big( 3 f_1 + f_2 \big)^2 (N - 1) \big) + (N -
  1) \mu^{- 2 \delta} \lambda_Y^2  \big( 3 (- 3 c_{6, 1} - 2 c_{6, 2} + c_3
  \mu^2 + 5 c_5) \nonumber\\&&+ \big( 3 f_1 + f_2 \big)^2 (N - 1) \big) + (3 f_1 -
  f_2) (N - 1) \mu^{- \delta} \lambda_Y + 3 (3 f_1 - f_2) \mu^{- \delta}
  \lambda_T + 1 \bigg) .
\end{eqnarray*}
We define the renormalized coupling constants to be
\begin{equation}
  g_Y = \mu^{- \delta} \Gamma_{\lambda_Y}, \quad g_T = \mu^{- \delta}\Gamma_{\lambda_T},
  \quad g_{\tmop{ST}} = \mu^{- (D - 2 \Delta_{\psi} - 1)} \Gamma_{\tmop{ST}},
  \quad \tmop{and} \quad g_{\tmop{Mass}} = \mu^{- (D - 2 \Delta_{\psi})}
  \Gamma_{\tmop{Mass}} . \label{3Dfermioncoupling} \quad
\end{equation}
By inverting the above relations, we get the beta functions
\[ \beta_Y = \mu \frac{\partial}{\partial \mu} g_Y, \quad \beta_T = \mu
   \frac{\partial}{\partial \mu} g_T, \quad \beta_{\tmop{ST}} = \mu
   \frac{\partial}{\partial \mu} g_{\tmop{ST}}, \quad \beta_{\tmop{Mass}} =
   \mu \frac{\partial}{\partial \mu} g_{\tmop{Mass}} \]
for $g_Y$, $g_T$, $g_{\tmop{ST}}$ and $g_{\tmop{Mass}}$, which then gives us
\eqref{3dfermionbeta}, \eqref{3dfermionT} and \eqref{3dfermionMass}.

When calculating $\Gamma_{\mathrm{ST}}$, we can again take the special kinetic point $q^\mu=0$ as the bosonic case. An argument similar to the scalar case tells us that one can not build an invariant tensor that transforms in the spin-2 representation of the Euclidean group (we can not use the external momentum since $q^\mu=0$.) This means only the second diagram in Fig. 10 contributes to the renormalization to $T_{\mu \nu}$.

For the $4+1$ Dermionic model, we can calculate the corresponding renormalization functions. The renormalization function for the four-fermion coupling is
\begin{eqnarray}
 \Gamma_{\alpha \beta \gamma \delta, a b c d} 
= && \Gamma_{\lambda_{T_2}}\bigg(\Omega_{a d} \Omega_{b c}\left(C \cdot \gamma^{\mu \nu}\right)_{\alpha \delta}\left(C \cdot \gamma_{\mu \nu}\right)_{\beta \gamma}+\Omega_{a b} \Omega_{c d}\left(C \cdot \gamma^{\mu \nu}\right)_{\alpha \beta}\left(C \cdot \gamma_{\mu \nu}\right)_{\gamma \delta}\nonumber\\&&+\Omega_{a c} \Omega_{b d}\left(C \cdot \gamma^{\mu \nu}\right)_{\alpha \gamma}\left(C \cdot \gamma_{\mu \nu}\right)_{\beta \delta}\bigg),
\end{eqnarray}
with
\begin{eqnarray*}
  \Gamma_{\lambda_{T_2}} & = & \lambda_{T_2} + (8 f_2 (N + 2) - 8 f_1 (N -
  28)) \mu^{- \delta} \lambda_{T_2}^2 \nonumber\\&& + 64 \mu^{- 2 \delta} \lambda_{T_2}^3
  \big( - 2 (59 N + 3) c_{6, 0} - 118 Nc_{6, 1} + 20 Nc_{6, 2} - 6 c_{6, 1} +
  20 c_{6, 2} + 58 c_3 \mu^2 N \nonumber\\&&+ 118 c_3 \mu^2 + c_5 (14 - 98 N) + f_1^2 N^2 +
  f_2^2 N^2 - 2 f_1 f_2 N^2 + 3 f_1^2 N + 3 f_2^2 N \nonumber\\&&- 6 f_1 f_2 N + 787 f_1^2
  + 35 f_2^2 - 6 f_1 f_2 \big) .
\end{eqnarray*}

The renormalization function of the stress tensor-fermion-fermion vertex (at $q^\mu=0$ ) is
with
$$
\begin{gathered}
\left(\Gamma^{\mu \nu}\right)_{\alpha \beta, a b}=2 \cdot i \cdot \Gamma_{\mathrm{ST}}\left(p^\nu \cdot\left(C \cdot \gamma^\mu\right)_{\alpha \beta}+p^\nu \cdot\left(C \cdot \gamma^\mu\right)_{\alpha \beta}-\frac{2}{5} \cdot \eta^{\mu \nu} p^\sigma \cdot\left(C \cdot \gamma^\sigma\right)_{\alpha \beta}\right) \Omega_{a b}, \\
\Gamma_{\mathrm{ST}}=\lambda_{\mathrm{ST}}+1920 \cdot c_5 \cdot(N+1) \cdot \mu^{-2 \delta} \cdot \lambda_{\mathrm{ST}} \cdot \lambda_{T_2}^2
\end{gathered}
$$
The renormalization function of mass-fermion-fermion vertex is,
\[ \Gamma_{\alpha \beta, a b} = \Gamma_{\tmop{Mass}} C_{\alpha \beta}
   \delta_{a b}, \]
with
\begin{eqnarray*}
  \Gamma_{{\rm Mass}} & = & \lambda_{{\rm Mass}}  (1 + 40 (5 f_1 - f_2) \mu^{-
  \delta} \lambda_{T_2} \nonumber\\&&+ 320 (N - 3) \mu^{- 2 \delta} \lambda_{T_2}^2  (- 5
  c_{6, 1} - 2 c_{6, 2} + 7 c_5 + c_3 \mu^2) + 1600 (5 f_1 + f_2)^2 \mu^{- 2
  \delta} \lambda_{T_2}^2) .
\end{eqnarray*}
We define the renormalized coupling constants to be
\[ g_{T_2} = \mu^{- \delta} \Gamma_{T_2}, \quad \quad g_{\tmop{ST}} = \mu^{-
   (D - 2 \Delta_{\psi} - 1)} \Gamma_{\tmop{ST}}, \quad \tmop{and} \quad
   g_{\tmop{Mass}} = \mu^{- (D - 2 \Delta_{\psi})} \Gamma_{\tmop{Mass}} .
   \quad \]
By inverting the above relations, and use we can get the beta functions
\[ \quad \beta_{T_2} = \mu \frac{\partial}{\partial \mu} g_{T_2}, \quad
   \beta_{\tmop{ST}} = \mu \frac{\partial}{\partial \mu} g_{\tmop{ST}}, \quad
   \beta_{\tmop{Mass}} = \mu \frac{\partial}{\partial \mu} g_{\tmop{Mass}} \]
for $g_{T_2}$, $g_{\tmop{ST}}$ and $g_{\tmop{Mass}}$, which then gives us
\eqref{5Dfermionbeta}, \eqref{5DfermionT} and \eqref{5DfermionMass}.

\bibliography{main.bib}
\bibliographystyle{utphys}

\end{document}